\documentclass[reprint,pra,longbibliography]{revtex4-1}

\usepackage{amsmath}
\usepackage{mathtools}
\usepackage{amssymb,amsfonts,bm}
\usepackage{dsfont}

\usepackage{graphicx}
\usepackage{float}
\usepackage{microtype}

\usepackage{braket}
\usepackage[separate-uncertainty]{siunitx}
\usepackage{hyphenat}

\usepackage[svgnames]{xcolor}
\usepackage[caption=false, labelfont=bf, labelformat=simple]{subfig}

\usepackage{hyperref}
\hypersetup{colorlinks,linkcolor={DarkBlue},citecolor={DarkBlue},urlcolor={DarkBlue}}
\usepackage[capitalize]{cleveref}

\usepackage{xcolor}

\renewcommand{\phi}{\varphi}
\renewcommand{\theta}{\vartheta}
\renewcommand{\epsilon}{\varepsilon}
\newcommand{\e}[0]{\text{e}} % euler constant
\newcommand{\ii}{\text{i}}
\newcommand{\Tr}[1]{\operatorname{Tr}{#1}} % trace operator
\newcommand{\C}{\mathbb{C}} % complex numbers
\newcommand{\R}{\mathbb{R}} % real numbers
\newcommand{\LandO}[1]{\mathcal{O}\left(#1\right)}
\newcommand{\diag}[1]{\operatorname{diag}{\left(#1\right)}}

\renewcommand{\openone}{\mathds{1}} % identity operator
\newcommand{\rank}[1]{\operatorname{rank}{#1}} % rank

\begin{document}

\title{Efficient construction of tensor-network representations of many-body Gaussian states}

\author{Alexander N{\"u}{\ss}eler}
\affiliation{Institut f{\"u}r Theoretische Physik and Center for Integrated Quantum Science and Technology (IQST),\\
Albert-Einstein-Allee 11, Universit{\"a}t Ulm, 89069 Ulm, Germany}

\author{Ish Dhand}
\altaffiliation[Present address: ]{Xanadu Quantum Technologies Inc., Toronto, ON, M5G 2C8, Canada}
\affiliation{Institut f{\"u}r Theoretische Physik and Center for Integrated Quantum Science and Technology (IQST),\\
Albert-Einstein-Allee 11, Universit{\"a}t Ulm, 89069 Ulm, Germany}

\author{Susana F.~Huelga}
\affiliation{Institut f{\"u}r Theoretische Physik and Center for Integrated Quantum Science and Technology (IQST),\\
Albert-Einstein-Allee 11, Universit{\"a}t Ulm, 89069 Ulm, Germany}

\author{Martin B.~Plenio}
\affiliation{Institut f{\"u}r Theoretische Physik and Center for Integrated Quantum Science and Technology (IQST),\\
Albert-Einstein-Allee 11, Universit{\"a}t Ulm, 89069 Ulm, Germany}

\begin{abstract}
We present a procedure to construct tensor-network representations of many-body Gaussian states efficiently and with a controllable error.
These states include the ground and thermal states of bosonic and fermionic quadratic Hamiltonians, which are essential in the study of quantum many-body systems. 
The procedure improves computational time requirements for constructing many-body Gaussian states by up to five orders of magnitude for reasonable parameter values, thus allowing simulations beyond the range of what was hitherto feasible.
Our procedure combines ideas from the theory of Gaussian quantum information with tensor-network based numerical methods thereby opening the possibility of exploiting the rich tool-kit of Gaussian methods in tensor-network simulations.
\end{abstract}

\maketitle

\textit{Introduction\textemdash}%
The study of quantum many-body systems is one of the key challenges in modern quantum physics.
While certain systems allow for analytical treatment, the vast majority requires numerical methods.
The regime of applicability of most numerical methods is constrained by the Hilbert-space dimension growing exponentially with the system size.
Fortunately, in many physically relevant applications tensor-networks have shown to defy this curse of dimensionality and have thus become an important tool for efficiently simulating bosonic as well as fermionic many-body systems.
However, the initialisation of such tensor-networks often remains computationally demanding.

In particular, tensor-networks have been applied extensively in the study of open quantum system dynamics and quantum thermodynamics in the condensed phase.
Important applications include the study of quantum impurity models such as the Kondo model~\cite{Weichselbaum2009}; the dynamics of populations and coherences in light induced processes in natural photosynthetic complexes~\cite{Chin2013,Prior2010,strathearn2018}; the time-frequency spectrum of the environmental excitations in spin-boson models~\cite{Schroeder2016}; polaron-polaritons in organic microcavities~\cite{DelPino2018}; quantum thermal machines comprising a system coupled to multiple fermionic baths~\cite{Brenes2019}.
Naturally all these problems take place at finite temperature.
Thus, when studying the dynamics of these systems, it is reasonable to assume that system and environment are initially prepared in a product state with the environment being in a thermal state with respect to its Hamiltonian~\cite{Bulla2008,Prior2010,Chin2010}.
This environment Hamiltonian is usually modelled as Gaussian, i.e., it only comprises terms that are quadratic in the ladder operators.
Hence, any time evolution involves the construction of a tensor-network representation of the Gaussian thermal states of the environment at first.

In order to prepare these initial thermal states of the environment, state-of-the-art methods rely on imaginary time evolution.
More precisely, these methods start by constructing a completely mixed state and successively cooling this state down to the desired temperature by propagating it incrementally in the inverse temperature. 
By the inherent design of such methods, imaginary time evolution becomes increasingly expensive with decreasing temperature as the total propagation length increases.
As a consequence, a tensor-network based simulation of many-body systems in the low and the intermediate temperature regime was hitherto infeasible.

For the specific application of simulating the dynamics of open quantum systems, alternative methods such as the thermofield-based or the thermalized TEDOPA approach~\cite{DeVega2015,Tamascelli2019,Nuesseler2019} have been proposed, which include the thermal dependence into the bath Hamiltonian by exploiting thermal Bogoliubov transformations.
However, it remains unclear under which circumstances these methods really decrease the computational requirements during the simulation of the dynamics.
Furthermore, thermalized TEDOPA is restricted to simulating interaction Hamiltonians that comprise only a single interaction term that factorizes between system and each environment.
Hence, even for the specialized case of simulating open systems, a general method to construct tensor-network representations of Gaussian states is still of great interest in order to efficiently simulate systems coupled to bosonic and fermionic environments.

Here, we present such a method by combining tools from Gaussian quantum information and quantum optics to construct Gaussian states using local thermal states, squeezers and passive linear optical circuits.
We show that this method provides substantial speed-up over existing methods in the low and the intermediate temperature regime.
To determine the speed-up we perform a thorough analysis of the number of basic floating-point operations (fpos) which gives a device- and implementation-independent estimate of the computational complexity.
Thus, due to the drastically reduced computational demands, our method enables the simulation of quantum many-body systems in hitherto inaccessible temperature regimes.

\textit{Procedure\textemdash}%
We now introduce our procedure, which takes a Gaussian Hamiltonian as input and yields an MPO representation of the thermal state at the desired inverse temperature $\beta$ as output.
In particular, we proceed in two steps.
First, the given quadratic Hamiltonian is brought into a diagonal form by an appropriately chosen Bogoliubov transformation.
The global thermal state of this set of non-interacting modes, called the normal modes, is a product state with each mode being in its local thermal state~\cite{Serafini2017a,Adesso2014}.
Analytically constructing the matrix-product operator representation of this product state completes the first step. 
Second, the unitary operator associated with the inverse Bogoliubov transformation is implemented efficiently using ideas from quantum optics.
In particular, we decompose the unitary operator into a circuit of local beam splitters, phase shifters and squeezers.
Hence, the thermal state of the interacting modes is ultimately obtained by applying this circuit on the MPO representation of the thermal states of the normal modes.

In more detail, we consider a general Gaussian Hamiltonian of $N$ modes defined by
\begin{align}
	\hspace{-6.5pt}\hat{H} &:= \sum_{i,j = 1}^{N} \left(\alpha_{ij} \hat{a}_i^\dagger \hat{a}_j + \nu \alpha_{ij}^* \hat{a}_i \hat{a}_j^\dagger + \zeta_{ij} \hat{a}_i \hat{a}_j + \nu \zeta_{ij}^* \hat{a}_i^\dagger \hat{a}_j^\dagger \right)
\end{align}
where $\nu = 1$ ($\nu = -1$) for bosons (fermions).
Furthermore, $\alpha = (\alpha_{ij})$ is hermitian and $\zeta = (\zeta_{ij})$ is symmetric (anti-symmetric) for bosons (fermions).
Collecting the bosonic (fermionic) ladder operators in a vector, $\hat{\bm{a}} := (\hat{a}_1, \dots, \hat{a}_N, \hat{a}_1^\dagger, \dots, \hat{a}_N^\dagger)^{\intercal}$, $\hat{\bm{a}}^\dagger := (\hat{a}_1^\dagger, \dots, \hat{a}_N^\dagger, \hat{a}_1, \dots, \hat{a}_N)$, the Hamiltonian $\hat{H}$ can be written as bilinear form $\hat{H} = \hat{\bm{a}}^\dagger H \hat{\bm{a}}$ with the Hamilton matrix
\begin{align}
	H &:= 
	\begin{pmatrix}
		\alpha & \nu\zeta^* \\
		\zeta  & \nu\alpha^*
	\end{pmatrix}\in \C^{2N \times 2N}.
	\label{Eq:general_hamilton_matrix}
\end{align}
Bosonic and fermionic operators naturally obey canonical (anti-) commutation relations (CCR/CAR) which can now be compactly expressed as
\begin{align}
	[\hat{\bm{a}}_i, \hat{\bm{a}}_j^\dagger]_{\pm} &= (\Omega_{\pm})_{ij},
	&
	\Omega_{\pm} &= 
	\begin{pmatrix}
		\openone & 0 \\
		0        & \nu \openone
	\end{pmatrix}.
	\label{Eq:general_commutation_relations}
\end{align}

Given a Hamilton matrix $H$, which we assume to be positive-definite in the bosonic case, there exists a Bogoliubov transformation $\hat{\bm{a}} \mapsto \hat{\bm{b}} := T \hat{\bm{a}}$ preserving \cref{Eq:general_commutation_relations} and
\begin{align}
	\hat{H} &= \hat{\bm{a}}^\dagger H \hat{\bm{a}} = \hat{\bm{b}}^\dagger (T^{-1})^\dagger H T^{-1} \hat{\bm{b}} = \hat{\bm{b}}^\dagger (D \oplus \nu D) \hat{\bm{b}}
	\label{Eq:general_normal_mode_decomposition}
\end{align}
where $D := \diag{d_1,\dots,d_N}$ and $d_i > 0 \, \forall i$.
This is called the normal mode decomposition of $\hat{H}$.
It can be shown that $T$ is the solution of a general eigenvalue problem and can thus, in practice, be computed by standard linear algebra solvers (see~\cref{Sec:bosonic_gaussian_hamiltonians,Sec:fermionic_gaussian_hamiltonians} for details).

In general, a thermal state at inverse temperature $\beta$ is defined by $\hat{\rho}(\beta) := \frac{1}{\mathcal{Z}}\e^{-\beta \hat{H}}$ where $\mathcal{Z}$ denotes the partition function.
However, in the Fock basis $\ket{n_i}_{\hat{\bm{b}}}$ corresponding to the normal modes $\hat{\bm{b}}$, this expression simplifies to a product state of the form
\begin{align}
	\hat{\rho}_{\hat{\bm{b}}}(\beta) &:= \mathcal{N} \bigotimes_{i = 1}^{N} \sum_{n_i} \e^{-2\beta d_i n_i} \ket{n_i}_{\hat{\bm{b}}} \bra{n_i}_{\hat{\bm{b}}}
	\label{Eq:general_normal_mode_thermal_state}
\end{align}
with an appropriately chosen normalization constant $\mathcal{N}$.
While in the fermionic case the sum over Fock states $n_i$ is finite, in the bosonic case the sum is infinite and has to be truncated for numerical purposes.
However, since $\beta > 0$, the probability amplitudes decay exponentially and for a given accuracy we can truncate the Fock space of each mode at some finite value $M-1$.

Ultimately, our aim is to construct MPO representations,
\begin{align}
	\hat{\rho} &= \sum_{\substack{m_1, \dots, m_N = 1\\ n_1, \dots, n_N = 1}}^{d} A_{m_1}^{n_1} A_{m_2}^{n_2} \cdots A_{m_N}^{n_N} \ket{\bm{m}} \bra{\bm{n}}
\end{align}
of thermal states of $\hat{H}$ in the Fock basis.
Here, $A^{n_i}_{m_i} \in \C^{r_{i-1} \times r_i}$ where the $r_i$ are referred to as bond dimensions and $r_0 = r_N = 1$.
Hence, from \cref{Eq:general_normal_mode_thermal_state} we obtain an MPO representation of $\hat{\rho}_{\hat{\bm{b}}}(\beta)$ with bond dimensions $r_i = 1 \; \forall i$ by setting $A^{n_i}_{m_i} := \e^{-2\beta d_i n_i} \delta_{n_i,m_i}$.
In order to obtain the thermal state $\hat{\rho}_{\hat{\bm{a}}}(\beta)$, we have to invert the normal mode transformation of \cref{Eq:general_normal_mode_decomposition}.
On the level of density operators, this inverse transformation $T^{-1}$ corresponds to a unitary generator $\hat{T}$ such that $\hat{\rho}_{\hat{\bm{a}}}(\beta) = \hat{T}^\dagger \hat{\rho}_{\hat{\bm{b}}}(\beta) \hat{T}$~\cite{Serafini2017a}.
In the following we will show how this unitary operator $\hat{T}$ can be efficiently implemented in terms of a tensor-network using ideas from quantum optics.

Gaussian unitary operators are categorized into active and passive transformations.
Passive transformations conserve the number of excitations and can be implemented by a network of beam splitters and phase shifters~\cite{Reck1994a,Clements2016a}. 
In contrast, active transformations do not conserve the number of excitations and therefore additionally require squeezing operations for their implementation.
More precisely, the Bloch-Messiah decomposition~\cite{Ring1980} allows decomposing any active transformation $T$ into a product of three matrices, $T = \bar{U} \bar{S} \bar{V}$, where $\bar{U}$ and $\bar{V}$ are passive transformations and $\bar{S}$ represents a squeezing transformation.

We propose to decompose the passive part of the transformation either as a triangular network of beam splitters and phase shifters of depth $2N-3$~\cite{Reck1994a} or a rectangular network of depth $N$~\cite{Clements2016a} both followed by an extra layer of phase shifters.
Here, each of the beam splitter elements is a two-mode gate of the form $\hat{B}(\theta) = \e^{\ii \theta (\hat{a}_i^\dagger \hat{a}_{i+1} + \nu \hat{a}_i \hat{a}_{i+1}^\dagger)}$ that admits an exact MPO representation with bond dimension $r_i = M^2,\, r_{j\ne i} = 1$.
The phase shifters are single-mode gates of the form $\hat{P}(\phi) = \e^{\ii \phi \hat{a}_i^\dagger \hat{a}_i}$ and can be represented by an MPO with bond dimension $r_i = 1\,\forall i$.
In practice, given the unitary matrix $\bar{U}$ of the passive transformation, the angles for the beam splitters and phase shifters can be obtained numerically using open-source libraries such as~\cite{Killoran2019a}.

The form and the implementation of the squeezing matrix $\bar{S}$ depends on the species of the particles.
The bosonic Bloch-Messiah theorem~\cite{Ring1980} states that $\bar{S}$ corresponds to a single mode squeezing operation with generator $\hat{S}(z_i) = \e^{z_i (\hat{a}_i^2 - (\hat{a}_i^\dagger)^2)}$, $z_i \in \R$.
The total squeezing layer is then formed by taking the tensor product of the single mode squeezers and admits again an MPO representation of bond dimension $r_i = 1\,\forall i$.
For fermions, the fermionic Bloch-Messiah theorem~\cite{Ring1980} states that $\bar{S}$ distinguishes between paired and blocked modes; for paired modes $\bar{S}$ is a two-mode squeezing operation with generator $\hat{S}(z_i) = \e^{z_i (\hat{a}_i^\dagger \hat{a}_{i+1}^\dagger - \hat{a}_{i} \hat{a}_{i+1})}$, $z_i \in [0,2\pi)$, and for blocked modes $\bar{S}$ is either the identity or the swap operation defined by $\hat{x}_i = \hat{a}_i + \hat{a}_i^\dagger$.
The complete squeezing layer is again assembled by taking the tensor product of the local operators and admits an MPO representation with $r_i \leq 4 \, \forall i$.

\begin{figure}
	\includegraphics[width=\linewidth]{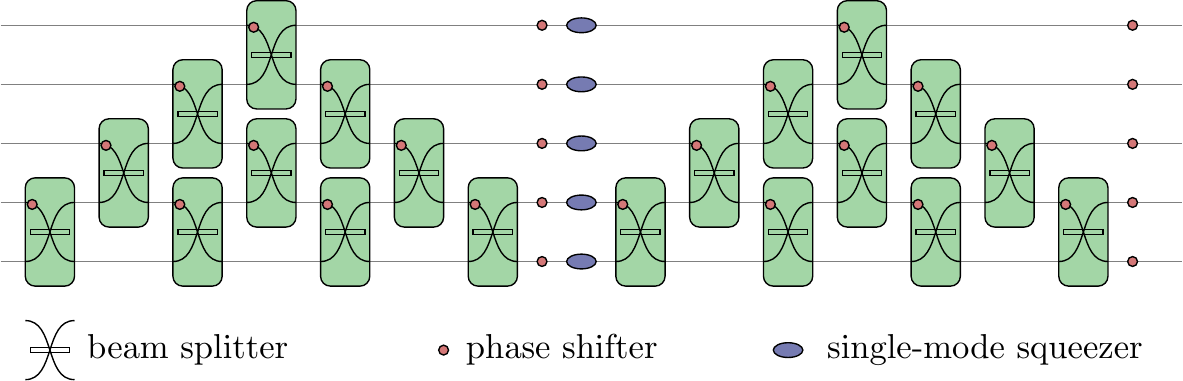}
	\caption{
	\textbf{Tensor-network implementation of an active, bosonic normal mode transformation:}
	The active bosonic normal mode transformation is split into two passive transformations sandwiching a local active transformation.
	Implementing the passive transformations according to Reck et al.~\cite{Reck1994a}, the initial product state first passes a triangular shaped sequence of phase-shifters and beam-splitters followed by another layer of single-mode phase-shifters.
	Subsequently, a layer of single-mode squeezing operators accounts for the active part of the transformation.
	After passing a second passive transformation we obtain the final, correlated thermal state.
	}
	\label{Fig:full_reck_circuit}
\end{figure}

While for bosons we can directly infer the MPO representations of the single and two-mode operations from its corresponding matrix representations, this is not possible for fermions due to the CAR.
This problem is addressed by mapping the fermionic ladder operators via the Jordan-Wigner transformation~\cite{Jordan1928,Nielsen2005,Parkinson2010} onto Pauli spin operators whose intrinsic algebra encodes the CAR.
From the matrix representation of the spin operators we are then able to derive the corresponding MPO decomposition, see~\cref{Sec:jordan_wigner_transformation}.
\cref{Fig:full_reck_circuit} exemplifies the tensor-network emerging from the considerations in this section.

\textit{Error sources\textemdash}%
The preparation scheme proposed in this work suffers from, mainly, two controllable sources of error~\textendash~one of which is specific to bosons.
While fermions are described by a finite dimensional Fock space, bosons live in an infinite dimensional Hilbert space.
In order to obtain a tensor-network representation we truncate the Hilbert space dimension to a finite number $M$.
This truncation can be performed in the Fock basis wherein only the lowest $M$ Fock basis states are considered.
Since states with lower energies are typically more likely to arise in dynamics than states with higher energies, this truncation error systematically reduces by increasing $M$.
A thorough analysis of the error introduced by this truncation is presented in \cite{Woods2015}.

Furthermore, within the preparation of the desired state we successively apply MPOs to the product state.
Without further action this will increase the bond dimensions of the tensor network representation exponentially in the number of applied gates, thereby leading to infeasible computational time and memory requirements.
To overcome this problem, we truncate the bond dimensions after each MPO-MPO product using a SVD compression scheme~\cite{Schollwock2011a}.
The error of this compression scales with the sum of the discarded singular values on each site which is, in turn, controlled by either constraining the maximal value of this sum or by fixing a certain number of singular values to retain.

Finally, we emphasize that unlike preparation schemes based on performing imaginary time evolution, our method involves neither an error due to the discretisation of time nor due to a Suzuki-Trotter splitting of the propagator.

\textit{Figures of merit\textemdash}%
To compare the quality and performance of our preparation scheme with the imaginary time evolution, we introduce two figures of merit.
The first figure captures how faithfully the state is prepared, i.e., what is the accuracy of the prepared state with respect to the exact state.
Any Gaussian state is uniquely determined by its first and second moments, $\bm{m}_{\beta, \hat{\bm{a}}} := (\Tr{[\hat{\rho}_{\beta} \hat{\bm{a}}_i]})_{i=1}^{N}$ and $\gamma_{\beta,\hat{\bm{a}}} := (\Tr{[\hat{\rho}_\beta \hat{\bm{a}}_{i}^\dagger \hat{\bm{a}}_j}])_{i,j=1}^{N}$, respectively. 
For thermal states these expressions can be evaluated analytically.
While the first moments vanish identically, $\bm{m}_{\beta, \hat{\bm{a}}}^{\text{ex}} = 0$, the second moments read $\gamma_{\beta,\hat{\bm{a}}}^{\text{ex}} = T^\dagger \left[\bar{n}_{\beta, \nu}(D) \oplus (\openone + \nu \bar{n}_{\beta, \nu}(D)) \right] T$ where $T$ and $D$ are defined according to \cref{Eq:general_normal_mode_decomposition} and $\bar{n}_{\beta, \nu} := (\exp{(2\beta D)} - \nu)^{-1}$.
Hence, we measure the accuracy of the prepared state by computing the absolute value of the first moments as well as the relative error of the second moments with respect to the Frobenius norm, i.e., $\epsilon_{\bm{m}} = \| \bm{m}_{\beta, \hat{\bm{a}}}\|_{\text{F}}$ and  $\epsilon_{\gamma}^{\text{rel}} := \| \gamma_{\beta,\hat{\bm{a}}} - \gamma^{\text{ex}}_{\beta,\hat{\bm{a}}} \|_{\text{F}} \|\gamma^{\text{ex}}_{\beta,\hat{\bm{a}}}\|_{\text{F}}^{-1}$, respectively.
More details on measuring the accuracy of the prepared state and specific examples can be found in~\cref{Sec:figures_of_merit}.

The second figure captures the computational resource cost of the schemes, which we quantify by estimating the number of floating point operations (fpos) required to prepare the state numerically.
Compared to CPU time this measure is independent of the hardware and the specific implementation.
The proposed method, as well as other standard time evolution schemes, e.g., Time-Evolving Block Decimation (TEBD)~\cite{Vidal2003,Vidal2004}, consist of a sequence of MPO-MPO products and successive MPO compressions.
While the complexity of the MPO-MPO product can be estimated straightforwardly, the number of fpos for the MPO compression depends strongly on the compression scheme employed.
While we focus on the standard SVD compression scheme here~\cite{Schollwock2011a}, the performance might further improve by using randomized SVD implementations~\cite{Halko2011, Tamascelli2015}.
A derivation of the complexity estimates is in~\cref{Sec:complexity_estimates}.
The actual fpo counts are then obtained by dynamically tracking the number of operations in each step of the algorithm.

\textit{Numerical examples\textemdash}%
Here we analyse the performance of our method compared to the standard method for thermal state construction, namely that of imaginary time evolution.
Although our procedure is not limited to Hamiltonians that comprise nearest-neighbour interactions, imaginary time evolution via TEBD~\cite{Vidal2004,Schollwock2011a} is tailored for such Hamiltonians.
Hence, for the purposes of comparison, we restrict ourselves to nearest-neighbour Hamiltonians.

\begin{figure}
	\includegraphics[width=\linewidth]{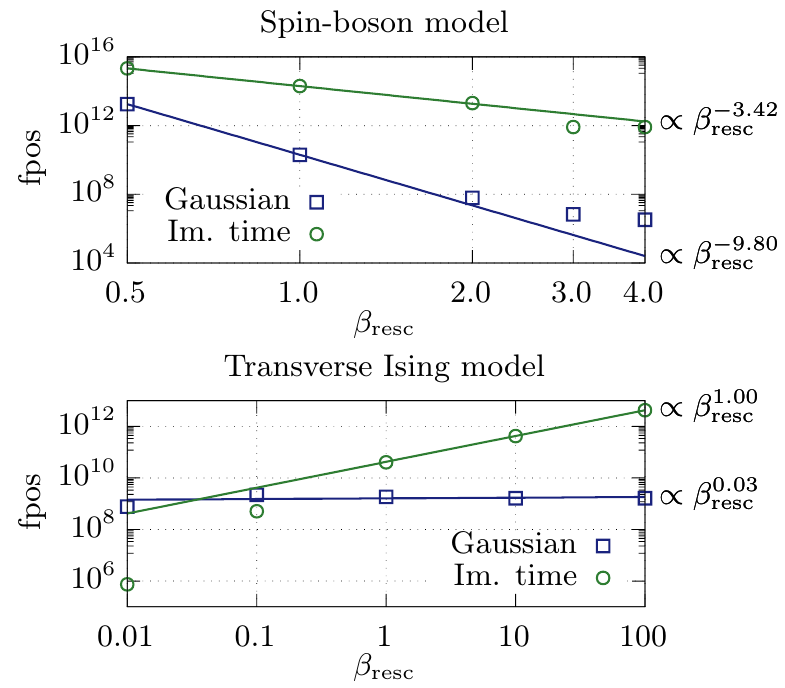}
	\caption{\textbf{Computational cost as a function of inverse temperature:}
	Number of floating point operations (fpos) required to prepare a thermal state at rescaled inverse temperature $\beta_{\text{resc}}$.
	For both models we fix $N = 20$ and demand $\epsilon_{\bm{m}}, \; \epsilon_{\gamma}^{\text{rel}} < 10^{-2}$.
	The remaining simulation parameters are chosen such that the number of fpos is minimized. 
	Straight lines correspond to non-linear least-squares fits to a power law.
	For further details we refer to \cref{Sec:numerical_examples}.
	}
	\label{Fig:fpos_vs_beta}
\end{figure}

As a first example, consider the prototypical spin-boson model with Ohmic spectral density~\cite{Weiss2012,Leggett1987}.
In particular, we aim to prepare a thermal state of the bath Hamiltonian
\begin{align}
	\hat{H} &= \sum_{i=1}^{N} \omega_i \hat{a}_i^\dagger \hat{a}_i + \sum_{i=1}^{N-1} t_i (\hat{a}_{i+1}^\dagger \hat{a}_{i} + \hat{a}_{i}^\dagger \hat{a}_{i+1}).
	\label{Eq:spin_boson_bath_hamiltonian}
\end{align}
Here, the frequencies $\omega_i$ and the nearest-neighbor couplings $t_i$ emerge from the TEDOPA chain mapping~\cite{Prior2010,Chin2010} with an Ohmic spectral density $J: [0,\lambda \omega_c] \rightarrow [0,\infty), \; J(\omega) :=  \omega  \e^{-\frac{\omega}{\omega_c}}$ where $\omega_c$ denotes the cut-off frequency and $\lambda$ is chosen such that $J(\lambda\omega_c)$ is below machine precision.
The top panel of \cref{Fig:fpos_vs_beta} depicts the number of fpos required to prepare the thermal state up to a given precision as a function of the inverse temperature.
More precisely, for the sake of comparability we rescale the inverse temperature $\beta$ by the energy gap between the ground and the first excited state and refer to this as $\beta_{\text{resc}}$.
We then fix upper bounds on $\epsilon_{\bm{m}}$ and $\epsilon_\gamma^{\text{rel}}$ and minimize the fpo count over the remaining simulation parameters.
See \cref{Sec:numerical_examples} for further details concerning this optimization.

\cref{Fig:fpos_vs_beta} shows that our proposed preparation scheme outperforms imaginary time evolution over a significant range of temperatures with the difference between both schemes becoming more pronounced as $\beta_{\text{resc}}$ increases, i.e., as we approach the low temperature regime.
This increasing advantage of our procedure can be explained as follows. 
For imaginary time evolution, larger $\beta_{\text{resc}}$ values require more steps to be performed.
Moreover, the required number of steps can scale faster than linearly in $\beta_{\text{resc}}$ since a longer evolution might also necessitate a decreased step size in order to keep the total Trotter error on a reasonable level.
In contrast, for our Gaussian preparation scheme, fewer excitations are present in the initial state for larger $\beta_{\text{resc}}$ values.
Thus, in general, fewer correlations build up in the circuit and the bond dimensions of the MPO stay lower.
This explains how the procedure introduced here offers a computational advantage of up to five orders of magnitude as compared to imaginary time evolution.

As a second example, we consider the transverse Ising model.
The Hamiltonian of this model reads
\begin{align}
	\hat{H} &= \sum_{i = 1}^N \hat{\sigma}_{i}^{z} + \lambda\; \sum_{i = 1}^{N-1} \hat{\sigma}_{i}^x \hat{\sigma}_{i+1}^{x},
\end{align}
where $\lambda$ denotes the ratio between the magnetic field strength in $z$-direction and the nearest-neighbor coupling strength in $x$-direction.
It is well known that this type of Hamiltonian can be mapped onto a fermionic Hamiltonian that is quadratic in the ladder operators via the Jordan-Wigner transformation~\cite{Jordan1928,Parkinson2010,Nielsen2005}.
In particular, the resulting Hamiltonian is not particle-preserving and thus the thermal state has to be obtained by an active Bogoliubov transformation.
The bottom panel of \cref{Fig:fpos_vs_beta} depicts the number of fpos required to prepare the thermal state as function of the inverse temperature for a fixed value of the coupling $\lambda = 1.2$.
Analogously to the previous example, we rescaled the inverse temperature by the energy gap, set upper bounds on $\epsilon_{\textbf{m}}$ and $\epsilon_{\gamma}^{\text{rel}}$ and optimized over the remaining simulation parameters (See \cref{Sec:numerical_examples}).
In contrast to the bosonic example, \cref{Fig:fpos_vs_beta} shows that for low $\beta_{\text{resc}}$ initialization by imaginary time evolution is preferable.
However, as $T$ decreases and $\beta_{\text{resc}}$ grows the computational cost increases significantly for imaginary time evolution, while it remains of the same order for our Gaussian preparation scheme across different temperature regimes. 

\begin{figure}
	\includegraphics[width=\linewidth]{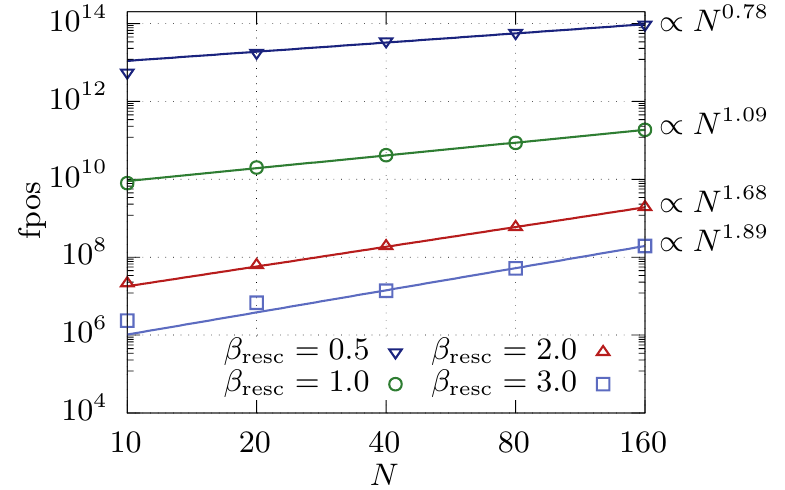}
	\caption{\textbf{Computational cost as a function of the number of modes:}
	Number of floating point operations (fpos) required to prepare a thermal state at rescaled inverse temperature $\beta_{\text{resc}}$ as a function of the number of modes $N$.
	Here, we demand relative errors $\epsilon_{\bm{m}}, \; \epsilon_{\gamma}^{\text{rel}} < 10^{-2}$ and optimize over the remaining simulation parameters.
	Straight lines correspond to non-linear least-squares fits of the data to a power law.
	For further details we refer to \cref{Sec:numerical_examples}.
	}
	\label{Fig:modes_vs_fpos}
\end{figure}

Another question is the scaling of the proposed method with the number of constituents $N$.
We address this question for the spin-boson model and various inverse temperatures $\beta_{\text{resc}}$ in \cref{Fig:modes_vs_fpos}.
Here, we fix again bounds on $\epsilon_{\bm{m}}$ and $\epsilon_{\gamma}^{\text{rel}}$ and optimize over the remaining simulation parameters (see~\cref{Sec:numerical_examples}).
Our results show that the proposed scheme scales polynomially in the number of modes with polynomial degree $\alpha < 2$ for all temperatures considered.
Moreover, we obtain that the polynomial order decreases as $\beta_{\text{resc}}$ increases.
Some intuition behind this scaling is presented in~\cref{Sec:supplementary_spin_boson_model}.
Note also that across the whole temperature regime the construction of a thermal state of $N=160$ modes with the proposed method requires less fpos than the construction of the corresponding thermal state of $N=20$ modes with imaginary time evolution, cf. \cref{Fig:fpos_vs_beta} and \cref{Fig:modes_vs_fpos}.

\textit{Conclusion\textemdash}%
In summary, we have successfully combined results from the fields of quantum optics and quantum information to obtain a highly efficient scheme to construct MPO representations of thermal states of Gaussian Hamiltonians.
This procedure applies to fermionic as well as bosonic Hamiltonians irrespective of whether they are excitation preserving or not.
Furthermore, our approach straightforwardly generalizes to arbitrary Gaussian states by complementing our method with an MPO representation of the displacement operator.
For ground states of excitation preserving, fermionic Hamiltonians our procedure reduces to the approach by Fishman et al.~\cite{fishman2015}.
Using the example of a spin-boson and a transverse Ising model, we have illustrated how our method outperforms standard imaginary time evolution via the TEBD algorithm significantly over a wide range of temperatures.
Hence, our method paves the way to explore temperature regimes that were hitherto inaccessible to tensor-network techniques. 

\section{Acknowledgements}%
We thank Myung-Joong Hwang and Mark Mitchison for helpful discussions.
Tensor-network simulations are performed using the \texttt{mpnum} package~\cite{Suess2017a} by Daniel Suess and Milan Holzaepfel.
This work is supported by the ERC Synergy grant HyperQ (Grant No 856432), the EU H2020 Quantum Technology Flagship project AsteriQs (Grant No 820394), the EU H2020 project Hyperdiamond (Grant No 667192), the BMBF (Federal Ministry of Education and Research) via NanoSpin and DiaPol, the German Research Foundation (DFG) via a Reinhart Koselleck project and grant no INST 40/467-1 FUGG (JUSTUS cluster) as well as the state of Baden-W{\"u}rttemberg through bwHPC.

\appendix

\newpage

\section{Bosonic Gaussian Hamiltonians}
\label{Sec:bosonic_gaussian_hamiltonians}

\subsection{Hamilton formulation}

Consider a bosonic quadratic Hamiltonian of the general form
\begin{align}
	\hat{H} &= \sum_{i,j = 1}^{N} \left(\alpha_{ij} \hat{a}_i^\dagger \hat{a}_j + \alpha_{ij}^* \hat{a}_i \hat{a}_j^\dagger + \zeta_{ij} \hat{a}_i \hat{a}_j + \zeta_{ij}^* \hat{a}_i^\dagger \hat{a}_j^\dagger \right)
	\label{Eq:general_bosonic_hamiltonian}
\end{align}
with $\alpha = (\alpha_{ij})$ hermitian and $\zeta = (\zeta_{ij})$ symmetric.
Here, $a_i^\dagger$ and $a_i$ denote bosonic creation and annihilation operators fulfilling the canonical commutation relations (CCR)
\begin{align}
[\hat{a}_i, \hat{a}_j] &= 0, &
[\hat{a}_i, \hat{a}_j^\dagger] &= \delta_{ij}, & 
&\forall i,j = 1,\dots,N.
\label{Eq:ccr_elements}
\end{align}
Defining the vectors of creation and annihilation operators
\begin{align}
	\hat{\bm{a}} &= (\hat{a}_1, \dots, \hat{a}_N, \hat{a}_1^\dagger, \dots, \hat{a}_N^\dagger)^{\intercal}, \\
	\hat{\bm{a}}^\dagger &= (\hat{a}_1^\dagger, \dots, \hat{a}_N^\dagger, \hat{a}_1, \dots, \hat{a}_N)
\end{align}
the Hamiltonian in \cref{Eq:general_bosonic_hamiltonian} can be expressed compactly as
\begin{align}
	\hat{H} &=
	\hat{\bm{a}}^\dagger
	H
	\hat{\bm{a}}
\end{align}
where
\begin{align}
	H &=
	\begin{pmatrix}
	\alpha & \zeta^* \\
	\zeta & \alpha^*
	\end{pmatrix}
	\label{Eq:bosonic_hamilton_matrix}
\end{align}
denotes the Hamilton matrix.
Exploiting the hermicity of $\alpha$ and the symmetry of $\zeta$ it follows that $H$ is hermitian.
Furthermore, rewriting the CCR of \cref{Eq:ccr_elements} in the vector representation leads to
\begin{align}
	[\hat{\bm{a}}_i, \hat{\bm{a}}_j^\dagger] &= \Omega_{ij} \quad \forall i,j = 1,\dots,N, \label{Eq:ccr_vector}
\end{align}
where
\begin{align}
	\Omega &:= 
	\begin{pmatrix}
		\openone & 0 \\
		0 & -\openone
	\end{pmatrix}.
\end{align}

\subsection{Bosonic Bogoliubov transformations}

In the following we are interested in linear transformations of the bosonic operators,
\begin{align}
	\hat{\bm{a}} \mapsto \hat{\bm{b}} = T \hat{\bm{a}},
	\label{Eq:linear_transformation_bosons}
\end{align}
induced by a matrix $T \in \C^{2N \times 2N}$.
In particular, the transformation and therefore the matrix $T$ has to preserve the algebraic structure of the bosonic ladder operators.
This kind of transformations is known as bosonic Bogoliubov transformations and is widely used in the literature~\cite{Schwabl2008}.
Any Bogoliubov transformation, admits the special block form
\begin{align}
	T &= 
	\begin{pmatrix}
		\gamma & \mu \\
		\mu^*  & \gamma^*
	\end{pmatrix}
	\label{Eq:bogoliubov_block_structure}
\end{align}
which ensures the adjointness of the transformed ladder operators.
Furthermore, we want the linear transformation to preserve the CCR.
Substituting \cref{Eq:linear_transformation_bosons} into \cref{Eq:ccr_vector} we find
\begin{align}
	[\hat{\bm{b}}_i, \hat{\bm{b}}_j^\dagger] &= \sum_{k,\ell} T_{i,k} T_{\ell,j}^{\dagger} [\hat{\bm{a}}_k, \hat{\bm{a}}_{\ell}^\dagger] \\
	&= \sum_{k,\ell} T_{i,k}  \Omega_{k,\ell} T_{\ell,j}^\dagger \overset{!}{=} \Omega_{i,j}.
\end{align}
Similarly, by looking at the commutator $[\hat{\bm{a}}_i^\dagger, \hat{\bm{a}}_j]$ we obtain
\begin{align}
	\sum_{k,\ell} T_{i,k}^\dagger \Omega T_{\ell,j} \overset{!}{=} \Omega_{ij}
\end{align}
and thus $T$ has to be such that 
\begin{align}
	T \Omega T^{\dagger} &= \Omega, & T^\dagger \Omega T   &= \Omega.
	\label{Eq:symplectic_relation}
\end{align}

In quantum optics, Bogoliubov transformations of the form of \cref{Eq:linear_transformation_bosons} with $\mu \not= 0$ are referred to as active transformations.
In contrast, transformations with $\mu = 0$ are referred to as passive transformations~\cite{Serafini2017a}.
This categorization will become particularly important in the actual implementation of the transformations in terms of optical circuits.

Bogoliubov transformations acting on the vector $\hat{\bm{b}}$ can also be related to unitary transformations of the bosonic operators itself.
In particular, for any bosonic Bogoliubov transformation defined by the matrix $T$ there exists a unitary operator $\hat{T}$ (also called the generator) such that~\cite{Serafini2017a}
\begin{align}
	\hat{b}_{i} &= (T\hat{\bm{a}})_i = \hat{T} \hat{a}_i \hat{T}^\dagger \quad \forall i = 1,\dots,N.
	\label{Eq:bosonic_bogoliubov_transformation_generator}
\end{align}

\subsection{Normal mode decomposition}

We will now show that for any positive-definite Hamilton matrix $H$, there exists a Bogoliubov transformation $T$ such that $\hat{\bm{b}} := T \hat{\bm{a}}$ and
\begin{align}
	\hat{H} &= \hat{\bm{b}}^\dagger (D \oplus D) \hat{\bm{b}}
\end{align}
where $D = \diag{d_1, \dots, d_N}$, $d_i > 0$ for all $i = 1,\dots,N$.

Consider the matrix $H^{\frac{1}{2}} \Omega H^{\frac{1}{2}}$ which is well-defined due to the positivity of $H$.
Furthermore, this matrix is hermitian and can thus be unitarily diagonalized such that~\cite{Serafini2017a}
\begin{align}
	U H^{\frac{1}{2}} \Omega H^{\frac{1}{2}} U^\dagger = D \oplus -D.
\end{align}
In the following we will show that the particular choice
\begin{align}
	T := (D^{-\frac{1}{2}} \oplus D^{-\frac{1}{2}}) U H^{\frac{1}{2}}
	\label{Eq:bosonic_bogoliubov_transformation_normal_modes}
\end{align}
is a valid Bogoliubov transformation and diagonalizes the Hamilton matrix $H$.
First, $T$ is again well-defined since $D$ is positive-definite and admits the block structure in \cref{Eq:bogoliubov_block_structure}.
Second, it holds
\begin{align}
	T \Omega T^\dagger &=  (D^{-\frac{1}{2}} \oplus D^{-\frac{1}{2}}) U H^{\frac{1}{2}} \Omega H^{\frac{1}{2}} U^\dagger (D^{-\frac{1}{2}} \oplus D^{-\frac{1}{2}}) \\
	&= (D^{-\frac{1}{2}} \oplus D^{-\frac{1}{2}}) (D \oplus -D) (D^{-\frac{1}{2}} \oplus D^{-\frac{1}{2}}) \\
	&= \Omega
\end{align}
as well as
\begin{align}
	T^\dagger \Omega T &= H^{\frac{1}{2}} U^\dagger (D^{-\frac{1}{2}} \oplus D^{-\frac{1}{2}}) \Omega (D^{-\frac{1}{2}} \oplus D^{-\frac{1}{2}}) U H^{\frac{1}{2}}\\
	&= H^{\frac{1}{2}} U^\dagger (D^{-1} \oplus -D^{-1}) U H^{\frac{1}{2}} \\
	&= \Omega.
\end{align}

Last but not least, $T$ diagonalizes the Hamilton matrix,
\begin{align}
	\begin{split}
		T^\dagger (D \oplus D) T &= H^{\frac{1}{2}} U^\dagger (D^{-\frac{1}{2}} \oplus D^{-\frac{1}{2}}) \\
		&\hspace{20pt}(D \oplus D) (D^{-\frac{1}{2}} \oplus D^{-\frac{1}{2}}) U H^{\frac{1}{2}}
	\end{split}\\
	&= H^{\frac{1}{2}} U^\dagger U H^{\frac{1}{2}} \\
	&= H.
\end{align}

\subsection{Excitation-preserving Hamiltonians}

In many physically relevant scenarios the Hamiltonian in \cref{Eq:general_bosonic_hamiltonian} is too general.
Consider for example the class of excitation preserving, quadratic Hamiltonians which follows from \cref{Eq:general_bosonic_hamiltonian} by setting $\zeta \equiv 0$.
The Hamiltonian thus admits a much simpler form,
\begin{align}
	\hat{H} &= \sum_{i,j = 1}^{N} \left(\alpha_{ij} \hat{b}_i^\dagger \hat{b}_j + \alpha_{ij}^* \hat{b}_i \hat{b}_j^\dagger \right),
\end{align}
and the Hamilton matrix in \cref{Eq:bosonic_hamilton_matrix} becomes block diagonal,
\begin{align}
	H &=
	\begin{pmatrix}
		\alpha & 0 \\
		0 & \alpha^*
	\end{pmatrix}.
\end{align}
For this sort of Hamiltonians the corresponding Gaussian theory simplifies significantly.

Since $\alpha$ is hermitian, there exists a unitary matrix $U$ such that $\alpha = U D U^\dagger$ and
\begin{align}
	T &:= 
	\begin{pmatrix}
		U & 0 \\
		0 & U^*
	\end{pmatrix}
\end{align}
defines a valid Bogoliubov transformation.
Furthermore, this $T$ transforms $H$ to its normal modes, i.e.,
\begin{align}
	\hat{H} &= \hat{\bm{b}}^\dagger (D \oplus D) \hat{\bm{b}}
\end{align}
with $\hat{\bm{b}} := T \hat{\bm{a}}$ and $D>0$.
Hence, an excitation-preserving Gaussian Hamiltonian can be diagonalized by applying a single passive transformation instead of a passive transformation followed by an active transformation followed by another passive transformation.

\subsection{Bosonic Bloch-Messiah decomposition}

The bosonic Bloch-Messiah decomposition states that any active Bogoliubov transformation can be decomposed into a product of three block matrices such that~\cite{Ring1980}
\begin{align}
	T &= \bar{U} \bar{S} \bar{V}^\dagger
\end{align}
where
\begin{align}
	\bar{U} &:= 
	\begin{pmatrix}
		U & 0 \\
		0 & U^*
	\end{pmatrix}, &
	\bar{S} &:= 
	\begin{pmatrix}
		S_\gamma & S_\mu \\
		S_\mu^*  & S_\gamma^*
	\end{pmatrix}, &
	\bar{V} &:= 
	\begin{pmatrix}
		V & 0 \\
		0 & V^*
	\end{pmatrix}.
\end{align}
Here, $U$ and $V$ are unitary and $S_\gamma$ and $S_\mu$ are diagonal.
Since $T$, $\bar{U}$ and $\bar{V}$ fulfill \cref{Eq:symplectic_relation} the same holds for $\bar{S}$.
Thus, the diagonal elements of $S_\gamma$ and $S_\mu$ fulfill
\begin{align}
	S_{\gamma,i}^2 - S_{\mu,i}^2 &= 1 \quad \forall i = 1,\dots, N.
\end{align}
Consequently there exist angles $\{\theta_k\}_{k=1}^N \subset \R$ such that
\begin{align}
	S_\gamma &= \operatorname{diag}{(\cosh{(\theta_1)},\dots,\cosh{(\theta_N)})} \\
	S_\mu    &= \operatorname{diag}{(\sinh{(\theta_1)},\dots,\sinh{(\theta_N)})}.
\end{align}
In quantum optics this corresponds to single mode squeezing applied to each of the $N$ modes.
Hence, any active Bogoliubov transformation $T$ can be implemented by a sequence of two passive transformations intercepted by a single mode squeezing operation. 

\subsection{Bosonic thermal states of Gaussian Hamiltonians}
\label{Sec:bosonic_thermal_states}

The thermal state of a Gaussian Hamiltonian at inverse temperature $\beta$ is defined by 
\begin{align}
	\hat{\rho}_{\beta} &:= \frac{\e^{-\beta \hat{H}}}{\mathcal{Z}}
	\label{Eq:def_bosonic_thermal_state}
\end{align}
where $\mathcal{Z} = \Tr{[\e^{-\beta \hat{H}}]}$ denotes the partition function.
In the following we will show that thermal states take a particularly simple form with respect to the normal modes.
Denoting the ladder operators of the normal modes by $\hat{\bm{b}}$ the partition function $\mathcal{Z}$ simplifies to
\begin{align}
	\mathcal{Z} &= \Tr{[e^{-\beta \hat{H}}]} \\
	&= \sum_{\bm{n}} \bra{\bm{n}} \e^{-\beta \sum_{i} d_i (a_i^\dagger a_i + a_i a_i^\dagger)} \ket{\bm{n}} \\
	&= \sum_{\bm{n}} \bra{\bm{n}} \e^{-\beta \sum_{i} d_i (2a_i^\dagger a_i + 1)} \ket{\bm{n}} \\
	&= \sum_{\bm{n}} \bra{\bm{n}} \e^{-\beta \sum_{i} d_i (2 n_i + 1)} \ket{\bm{n}}\\
	&= \prod_{i = 1}^{N} \e^{-\beta d_i} \sum_{n} \e^{-2\beta d_i n} \\
	&= \prod_{i = 1}^{N} \frac{\e^{-\beta d_i}}{1-\e^{-2\beta d_i}}.
	\label{Eq:bosonic_partition_function}
\end{align}
Substituting \cref{Eq:bosonic_partition_function} into \cref{Eq:def_bosonic_thermal_state}, we obtain
\begin{align}
	\hat{\rho}_{\beta} &= \frac{\e^{-\beta \hat{H}}}{\mathcal{Z}} \\
	&= \left(\prod_{i = 1}^{N} \frac{1 - \e^{-2\beta d_i}}{\e^{-\beta d_i}}\right) \e^{-\beta \sum_{j} d_j (2\hat{b}_j^\dagger \hat{b}_j + 1)} \\
	&= \prod_{i = 1}^{N} (1-\e^{-2\beta d_i}) \; \e^{-2\beta d_i \hat{b}_i^\dagger \hat{b}_i}.
	\label{Eq:bosonic_thermal_state}
\end{align}
Equivalently, $\hat{\rho}_{\beta}$ can be expressed in the corresponding Fock basis $\{\ket{\bm{n}}_{\hat{\bm{b}}}\}$ which yields
\begin{align}
	\hat{\rho}_{\beta, \hat{\bm{b}}} &= \left[\prod_{i = 1}^{N} (1-\e^{-2\beta d_i})\right] \; \left[ \bigotimes_{i = 1}^{N} \sum_{n_i = 0}^{\infty} \e^{-2\beta d_i n_i} \ket{n_i}_{\hat{\bm{b}}} \bra{n_i}_{\hat{\bm{b}}} \right].
	\label{Eq:bosonic_normal_mode_thermal_state_fock_basis}
\end{align}
Hence, $\hat{\rho}_{\beta, \hat{\bm{b}}}$ is a product state in the normal mode basis and admits an MPO representation with bond dimensions equal to 1.
The thermal state with respect to the initial modes $\hat{\bm{a}}$ is ultimately obtained by applying the inverse normal mode transformation.
Due to \cref{Eq:bosonic_bogoliubov_transformation_generator} we find
\begin{align}
	\hat{\rho}_{\beta, \hat{\bm{a}}} &= \hat{T}^\dagger \hat{\rho}_{\beta, \hat{\bm{b}}} \hat{T}.
\end{align}

Since $\hat{\rho}_{\beta}$ is a Gaussian state we can alternatively define it by its first and second moments. 
The first and second moments admit again a particularly simple form with respect to the normal mode basis $\hat{\bm{b}}$.
Henceforth, we collect the first moments in the vector $\bm{m}$ with entries
\begin{align}
	(\bm{m}_{\beta, \hat{\bm{b}}})_i &:= \Tr{[\hat{\rho}_{\beta} \hat{\bm{b}}_i]} \quad i = 1, \dots, 2N.
\end{align}
and the second moments in the covariance matrix with entries
\begin{align}
	(\gamma_{\beta, \hat{\bm{b}}})_{ij} &:= \Tr{[\hat{\rho}_{\beta} \hat{\bm{b}}_i^\dagger \hat{\bm{b}}_j]} \quad i,j = 1,\dots,2N.
\end{align}

Exploiting \cref{Eq:bosonic_thermal_state}, $\bm{m}_{\beta,\hat{\bm{b}}}$ as well $\gamma_{\beta, \hat{\bm{b}}}$ can be further simplified.
Firstly, it follows that the first moments of $\hat{\rho}_{\beta}$ vanish, i.e., 
\begin{align}
	\Tr{[\hat{\rho}_{\beta} \hat{b}_i]} = \Tr{[\hat{\rho}_{\beta} \hat{b}_i^\dagger]} = 0, \quad i = 1,\dots,N.
\end{align}
This means that the vector of first moments, $\bm{m}_{\beta, \hat{\bm{b}}}$, vanishes identically for any inverse temperature $\beta$.
Secondly, the second moments of $\hat{\rho}_{\beta}$ read
\begin{align}
	\Tr{[\hat{\rho}_{\beta} \hat{b}_i \hat{b}_j]} &= \Tr{[\hat{\rho}_{\beta} \hat{b}_i^\dagger \hat{b}_j^\dagger]} = 0
\end{align}
as well as
\begin{align}
	\Tr{[\hat{\rho}_{\beta} \hat{b}_i^\dagger \hat{b}_j]} &= \Tr{\left[\prod_{\ell} (1-\e^{-2\beta d_\ell}) \; \e^{-2\beta d_\ell \hat{a}_\ell^\dagger \hat{a}_\ell} \; \hat{b}_i^\dagger \hat{b}_j \right]} \\
	&= \delta_{ij} (1 - \e^{-2\beta d_i}) \sum_{n_i} \e^{-2\beta d_i n_i} \; n_i \\
	&= \delta_{ij} (1 - \e^{-2\beta d_i}) \frac{\e^{-2\beta d_i}}{(1 - \e^{-2\beta d_i})^2} \\
	&= \delta_{ij} \frac{1}{\e^{2\beta d_i} - 1} 
\end{align}
and
\begin{align}
	\Tr{[\hat{\rho}_{\beta} \hat{b}_i \hat{b}_j^\dagger]} &= \Tr{\left[\hat{\rho}_{\beta} (\delta_{ij} + \hat{b}_j^\dagger \hat{b}_i)\right]}\\
	&= \delta_{ij}\Tr{[\hat{\rho}_{\beta}]} + \Tr{[\hat{\rho}_{\beta} \hat{b}_j^\dagger \hat{b}_i]}  \\
	&= \delta_{ij} \left( 1 + \frac{1}{\e^{2\beta d_i} - 1}\right) \\
	&= \delta_{ij} \frac{\e^{2\beta d_i}}{e^{2\beta d_i} - 1}
\end{align}
for all $i,j = 1,\dots,N$.
Thus, the covariance matrix with respect to the normal modes reads
\begin{align}
	\gamma_{\beta, \hat{\bm{b}}} = \frac{1}{\e^{2\beta D} - 1} \oplus \frac{\e^{2\beta D}}{\e^{2\beta D} - 1}.
\end{align}

In order to obtain the covariance matrix in the original basis we apply the inverse bosonic Bogoliubov transformation of \cref{Eq:bosonic_bogoliubov_transformation_normal_modes}, i.e.,
\begin{align}
	\gamma_{\beta, \hat{\bm{a}}} &= T^\dagger \gamma_{\beta, \hat{\bm{b}}} T.
\end{align}

\section{Fermionic Gaussian Hamiltonians}
\label{Sec:fermionic_gaussian_hamiltonians}

\subsection{Hamilton formulation}

Consider a fermionic quadratic Hamiltonian of the general form
\begin{align}
	\hat{H} &= \sum_{i,j = 1}^{N} \left(\alpha_{ij} \hat{f}_i^\dagger \hat{f}_j - \alpha_{ij}^* \hat{f}_i \hat{f}_j^\dagger + \zeta_{ij} \hat{f}_i \hat{f}_j - \zeta_{ij}^* \hat{f}_i^\dagger \hat{f}_j^\dagger \right)
	\label{Eq:general_fermionic_hamiltonian}
\end{align}
with $\alpha = (\alpha_{ij})$ hermitian and $\zeta = (\zeta_{ij})$ antisymmetric.
Here, $f_i^\dagger$ and $f_i$ denote fermionic creation and annihilation operators fulfilling the canonical anti-commutation relation (CAR)
\begin{align}
\{\hat{f}_i, \hat{f}_j\} &= 0, &
\{\hat{f}_i, \hat{f}_j^\dagger \} &= \delta_{ij}, & 
&\forall i,j = 1,\dots,N.
\label{Eq:car_elements}
\end{align}
Defining the vectors of creation and annihilation operators
\begin{align}
	\hat{\bm{f}} &= (\hat{f}_1, \dots, \hat{f}_N, \hat{f}_1^\dagger, \dots, \hat{f}_N^\dagger)^{\intercal} \\
	\hat{\bm{f}}^\dagger &= (\hat{f}_1^\dagger, \dots, \hat{f}_N^\dagger, \hat{f}_1, \dots, \hat{f}_N)
\end{align}
the Hamiltonian in \cref{Eq:general_fermionic_hamiltonian} can be expressed compactly as
\begin{align}
	\hat{H} &=
	\hat{\bm{f}}^\dagger
	H
	\hat{\bm{f}}
\end{align}
where
\begin{align}
	H &=
	\begin{pmatrix}
	\alpha & -\zeta^* \\
	\zeta & -\alpha^*
	\end{pmatrix}
	\label{Eq:fermionic_hamilton_matrix}
\end{align}
denotes the Hamilton matrix.
Due to the hermicity of $\alpha$ and the antisymmetry of $\zeta$, the Hamilton matrix is hermitian.
Moreover, in this notation the CAR read
\begin{align}
	\{\hat{\bm{f}}_i, \hat{\bm{f}}_j^\dagger\} = \delta_{ij} \quad i,j = 1,\dots,2N.
	\label{Eq:car_vector}
\end{align}

\subsection{Fermionic Bogoliubov transformation}

We are now interested in linear transformations of the fermionic operators,
\begin{align}
	\hat{\bm{f}} \mapsto \hat{\bm{g}} = T \hat{\bm{f}},
	\label{Eq:fermionic_transformation_ladder_operator}
\end{align}
that preserve the fermionic nature of the operators.
The preservation of the fermionic nature restricts the set of admissable matrices $T \in \C^{2N \times 2N}$ in two ways.
Firstly, $T$ has to be of special block structure
\begin{align}
	T &= 
	\begin{pmatrix}
		\gamma & \mu \\
		\mu^* & \gamma^*
	\end{pmatrix}.
	\label{Eq:fermionic_bogoliubov_block_structure}
\end{align}
to ensure the adjointness of the transformed ladder operators. 
Secondly, the transformed operators $\hat{\bm{g}}$ still have to fulfill the CAR in \cref{Eq:car_vector}.
In particular, substituting \cref{Eq:fermionic_transformation_ladder_operator} into \cref{Eq:car_vector}  we find
\begin{align}
	\{\hat{\bm{g}}_i,\hat{\bm{g}}^{\dagger}_j\} &= \sum_{k,\ell = 1}^{2N} T_{ik} (T^\dagger)_{\ell j}\{\hat{\bm{f}}_k, \hat{\bm{f}}_\ell^\dagger\} \\
	&= \sum_{k,\ell = 1}^{2N} T_{ik} (T^\dagger)_{\ell j} \delta_{k \ell} \overset{!}{=} \delta_{ij}
\end{align}
for all $i,j = 1,\dots,2N$.
Analogously, we obtain
\begin{align}
	\{\hat{\bm{g}}_i^\dagger,\hat{\bm{g}}_j\} &= \sum_{k,\ell = 1}^{2N} (T^\dagger)_{ik} T_{\ell j} \delta_{k \ell} \overset{!}{=} \delta_{ij}.
\end{align}
Hence, in contrast to bosons where $T$ has to fulfill the constraints in \cref{Eq:symplectic_relation}, for fermions $T$ just has to be unitary, i.e.,
\begin{align}
	T^\dagger T &= T T^\dagger = \openone.
\end{align}
Moreover, for each fermionic Bogoliubov transformation induced by such a matrix $T$ there exists a unitary operator $\hat{T}$ such that on the level of individual modes we have~\cite{Serafini2017a}
\begin{align}
	\hat{T} \hat{\bm{f}}_k \hat{T}^\dagger = (T\hat{\bm{f}})_k
	\label{Eq:generator_unitary_transformation_fermions}
\end{align}
for all $k = 1,\dots,2N$.

In analogy to bosons, it turns out to be convenient to distinguish between two categories of Bogoliubov transformations.
Given a transformation matrix of the form in \cref{Eq:fermionic_bogoliubov_block_structure} we call a transformation active if $\mu \not= 0$ and passive if $\mu = 0$.

\subsection{Normal mode decomposition}

Equipped with the notion of fermionic Bogoliubov transformations we aim to introduce the normal mode decomposition of a fermionic Gaussian Hamiltonian.
Since $H$ admits the block structure in \cref{Eq:fermionic_hamilton_matrix} and is hermitian, we find a unitary matrix $T$ of the form defined in \cref{Eq:fermionic_bogoliubov_block_structure} such that $\hat{\bm{g}} = T\hat{\bm{f}}$ and
\begin{align}
	H &= \hat{\bm{g}}^\dagger (D \oplus -D) \hat{\bm{g}}
	\label{Eq:fermionic_normal_mode_decomposition}
\end{align}
with $D = \diag{d_1,\dots,d_N}$, $d_i \in \R$ ~\cite{Nielsen2005}.

\subsection{Particle-preserving fermionic Hamiltonians}

In a wide range of physical applications the fermionic Hamiltonian under consideration is particle-conserving, i.e., $\zeta \equiv 0$.
In this case, a normal decomposition is achieved by a block-diagonal transformation
\begin{align}
	T &= 
	\begin{pmatrix}
		\eta & 0 \\
		0 & \eta^*
	\end{pmatrix}
\end{align}
which effectively means that creation and annihilation operators are only mixed among themselves.

\subsection{Fermionic Bloch-Messiah decomposition}

The fermionic Bloch-Messiah decomposition states that any active fermionic Bogoliubov transformation can be decomposed into a product of three matrices,
\begin{align}
	T &= \bar{U} \bar{S} \bar{V}^\dagger,
\end{align}
where $\bar{U}$ and $\bar{V}$ are passive and $\bar{S}$ is an active transformation~\cite{Ring1980}.
In particular, these matrices admit the block structure
\begin{align}
	\bar{U} &:= 
	\begin{pmatrix}
		U & 0 \\
		0 & U^*
	\end{pmatrix}, &
	\bar{S} &:= 
	\begin{pmatrix}
		S_\gamma & S_\mu \\
		S_\mu^*  & S_\gamma^*
	\end{pmatrix}, &
	\bar{V} &:= 
	\begin{pmatrix}
		V & 0 \\
		0 & V^*
	\end{pmatrix}
\end{align}
with $U$, $V$ unitary.
The active blocks $S_\gamma$ and $S_\mu$ itself admit the block diagonal form
\begin{align}
	S_\gamma &:= 
	\begin{pmatrix}
		1 &        &   &          &        &           &   &       &  \\
		  & \ddots &   &          &        &           &   &       &  \\
		  &        & 1 &          &        &           &   &       &  \\
		  &        &   & \Gamma_1 &        &           &   &       &  \\
		  &        &   &          & \ddots &           &   &       &  \\
		  &        &   &          &        & \Gamma_N  &   &       &  \\
		  &        &   &          &        &           & 0 &       &  \\
		  &        &   &          &        &           &   &\ddots &  \\
		  &        &   &          &        &           &   &       & 0
	\end{pmatrix}
	\label{Eq:squeezing_matrix_1}\\
	S_\mu &:= 
	\begin{pmatrix}
		0 &        &   &     &        &      &   &       &  \\
		  & \ddots &   &     &        &      &   &       &  \\
		  &        & 0 &     &        &      &   &       &  \\
		  &        &   & \textrm{M}_1 &        &      &   &       &  \\
		  &        &   &     & \ddots &      &   &       &  \\
		  &        &   &     &        & \textrm{M}_N  &   &       &  \\
		  &        &   &     &        &      & 1 &       &  \\
		  &        &   &     &        &      &   &\ddots &  \\
		  &        &   &     &        &      &   &       & 1
	\end{pmatrix}\\
	\label{Eq:squeezing_matrix_2}
\end{align}
where the $2 \times 2$ blocks $\Gamma_k$ and $\textrm{M}_k$ are given by
\begin{align}
	\Gamma_k &:= 
	\begin{pmatrix}
		\gamma_k & 0 \\
		0        & \gamma_k
	\end{pmatrix}, \quad
	\textrm{M}_k := 
	\begin{pmatrix}
		0     & -\mu_k \\
		\mu_k & 0
	\end{pmatrix},
\end{align}
with $\gamma_k, \mu_k \in \R$.
Since $T$, $\bar{U}$ and $\bar{V}$ are unitary, $\bar{S}$ is also unitary.
On the level of $\Gamma_k$ and $\textrm{M}_k$ this translates to the property
\begin{align}
	\gamma_k^2 + \mu_k^2 = 1. 	
\end{align} 
Hence, $\gamma_k$ and $\mu_k$ can be parametrized in terms of an angle $\theta_k$ such that 
\begin{align}
	\gamma_k &= \cos{(\theta_k)}\\
	\mu_k    &= \sin{(\theta_k)}.
	\label{Eq:squeezing_angles_fermions}
\end{align}

In summary, looking at \cref{Eq:squeezing_matrix_1,Eq:squeezing_matrix_2} carefully we see that the transformation $\bar{S}$ distinguishes between two types of modes~\textemdash~paired and blocked modes.
For paired modes it acts like two-mode squeezing,
\begin{align}
	\begin{split}
		\hat{g}_k     &= \gamma_k \hat{f}_k - \mu_k \hat{f}_{k+1}^\dagger \\
		\hat{g}_{k+1} &= \gamma_k \hat{f}_{k+1} + \mu_k \hat{f}_k^\dagger
	\end{split}
\end{align}
whose generator is
\begin{align}
	\hat{S}(\theta_k) &= \e^{\theta_k (\hat{f}_k^\dagger \hat{f}_{k+1}^\dagger - \hat{f}_k \hat{f}_{k+1})}
\end{align}
with $\theta_k$ determined by \cref{Eq:squeezing_angles_fermions}.
For blocked modes $\bar{S}$ either acts as the identity (top left corner of $S_\gamma$ and $S_\mu$) or it performs a swap (bottom right corner of $S_\gamma$ and $S_\mu$).
The swap operation
\begin{align}
	\hat{g}_k &= \hat{f}_k^\dagger, \; \hat{g}_k^\dagger = \hat{f}_k
\end{align}
can be implemented easily by the operator $\hat{x}_i = \hat{f}_i + \hat{f}_i^\dagger$ since
\begin{align}
	\hat{g_k} &= \hat{x}_i \hat{f}_k \hat{x}_k = (\hat{f}_k + \hat{f}_k^\dagger) \hat{f}_k (\hat{f}_k + \hat{f}_k^\dagger) = (\hat{f}_k + \hat{f}_k^\dagger) \hat{f}_k \hat{f}_k^\dagger \nonumber \\
	&= \hat{f}_k^\dagger \hat{f}_k \hat{f}_k^\dagger = \hat{f}_k^\dagger ( 1 - \hat{f}_k^\dagger \hat{f}_k) \nonumber \\
	&= \hat{f}_k^\dagger
\end{align}
as well as 
\begin{align}
	\hat{g}_k^\dagger &= \hat{x}_k \hat{f}_k^\dagger \hat{x}_k = (\hat{f}_k + \hat{f}_k^\dagger) \hat{f}_k^\dagger (\hat{f}_k + \hat{f}_k^\dagger) = (\hat{f}_k + \hat{f}_k^\dagger) \hat{f}_k^\dagger \hat{f}_k \nonumber \\
	&= \hat{f}_k \hat{f}_k^\dagger \hat{f}_k = \hat{f}_k ( 1 - \hat{f}_k \hat{f}_k^\dagger) \nonumber \\
	&= \hat{f}_k.
\end{align}

\subsection{Fermionic thermal states of Gaussian Hamiltonians}

A general thermal state of a fermionic Gaussian Hamiltonian at inverse temperature $\beta$ and with chemical potential $\mu$ is defined by
\begin{align}
	\hat{\rho}(\beta) &= \frac{\e^{-\beta (\hat{H} - \mu \hat{N})}}{\mathcal{Z}},
	\label{Eq:general_fermionic_thermal_state}
\end{align}
where $\hat{N}$ denotes the global number operator.
For the sake of brevity we consider first the case of a vanishing chemical potential, i.e., $\mu = 0$.
In the end we generalize our findings to non-zero chemical potential.
As in the case of bosons, \cref{Eq:general_fermionic_thermal_state} can be evaluated further, however, with a fundamentally different result.
Exploiting the normal mode decomposition of $\hat{H}$ the partition function simplifies to
\begin{align}
	\mathcal{Z} &= \Tr{e^{-\beta \hat{H}}} \\
	&= \sum_{\bm{n}} \bra{\bm{n}} \e^{-\beta \sum_{i} d_i (\hat{g}_i^\dagger \hat{g}_i - \hat{g}_i \hat{g}_i^\dagger)} \ket{\bm{n}} \\
	&= \sum_{\bm{n}} \bra{\bm{n}} \e^{-\beta \sum_{i} d_i (2\hat{g}_i^\dagger \hat{g}_i - 1)} \ket{\bm{n}} \\
	&= \sum_{\bm{n}} \bra{\bm{n}} \e^{-\beta \sum_{i} d_i (2 n_i - 1)} \ket{\bm{n}}\\
	&= \prod_{i = 1}^{N} \e^{\beta d_i} \sum_{n=0,1} \e^{-2\beta d_i n} \\
	&= \prod_{i = 1}^{N} \e^{\beta d_i} \left(1 + \e^{-2\beta d_i}\right) \\
	&= \prod_{i = 1}^{N} \left( \e^{\beta d_i} + \e^{-\beta d_i} \right)
	\label{Eq:fermionic_partition_function}
\end{align}
Substituting this expression into the definition of the thermal state we obtain
\begin{align}
	\hat{\rho}_{\beta,\hat{\bm{g}}} &= \frac{\e^{-\beta \hat{H}}}{\mathcal{Z}} \\
	&= \left(\prod_{i = 1}^{N} \frac{1}{\e^{\beta d_i} + \e^{-\beta d_i}}\right) \e^{-\beta \sum_{j} d_j (2 \hat{g}_j^\dagger \hat{g}_j - 1)} \\
	&= \prod_{i = 1}^{N} \left(\frac{\e^{\beta d_i}}{\e^{\beta d_i} + \e^{-\beta d_i}}\right) \e^{-2\beta d_i \hat{g}_i^\dagger \hat{g}_i} \\
	&= \prod_{i = 1}^{N} \left(\frac{1}{\e^{-2\beta d_i} + 1}\right) \e^{-2\beta d_i \hat{g}_i^\dagger \hat{g}_i} .
	\label{Eq:fermionic_thermal_state}
\end{align}
In the Fock basis corresponding to the normal modes this is equivalent to
\begin{align}
	\hat{\rho}_{\beta, \hat{\bm{g}}} &= \left[\prod_{i = 1}^{N} \frac{1}{\e^{-2\beta d_i} + 1} \right] \left[\bigotimes_{i = 1}^{N} \sum_{n_i = 0,1} \e^{-2\beta d_i n_i} \ket{n_i}_{\hat{\bm{g}}}\bra{n_i}_{\hat{\bm{g}}} \right].
\end{align}
The thermal state in the original modes is then obtained by inverting the normal mode transformation in \cref{Eq:fermionic_normal_mode_decomposition}.
Due to \cref{Eq:generator_unitary_transformation_fermions} this translates to
\begin{align}
	\hat{\rho}_{\beta, \hat{\bm{f}}} &= \hat{T}^\dagger \hat{\rho}_{\beta,\hat{\bm{g}}} \hat{T}.
\end{align}

Since the fermionic thermal state is a Gaussian state, it can alternatively be defined in terms of the first and second moments of the creation and annihilation operators.
As we will see in the following, the normal mode basis proves most convenient to derive explicit expressions for the first and second moments. 
We collect the first moments in the vector
\begin{align}
	(\bm{m}_{\beta, \hat{\bm{g}}})_{i} &:= \Tr{[\hat{\rho}_\beta \hat{\bm{g}}_i]}, \quad i = 1,\dots, 2N
\end{align}
and the second moments in the covariance matrix
\begin{align}
	(\gamma_{\beta, \hat{\bm{g}}})_{ij} &:= \Tr{[\hat{\rho}_\beta \hat{\bm{g}}_i^\dagger \hat{\bm{g}}_j]}, \quad i,j = 1,\dots,2N.
\end{align}
Starting from \cref{Eq:fermionic_thermal_state} it follows firstly that the first moments of a thermal state at arbitrary inverse temperature $\beta$ vanish identically, i.e.,
\begin{align}
	\Tr{[\hat{\rho}_{\beta} \hat{g}_i]} &= \Tr{[\hat{\rho}_\beta \hat{g}_i^\dagger]} = 0 \quad i = 1, \dots, N.
\end{align}
Secondly, we obtain for the second moments
\begin{align}
	\Tr{[\hat{\rho}_{\beta} \hat{g}_i \hat{g}_j]} &= \Tr{[\hat{\rho}_{\beta} \hat{g}^\dagger \hat{g}_j^\dagger]} = 0
\end{align}
as well as
\begin{align}
	\Tr{[\hat{\rho}_{\beta} \hat{g}_i^\dagger \hat{g}_j]} &= \Tr{\left[\prod_{\ell=1}^{N} \frac{1}{\e^{-2\beta d_\ell} + 1} \e^{-2\beta d_\ell \hat{g}_\ell^\dagger \hat{g}_\ell} \; \hat{g}_i^\dagger \hat{g}_j \right]} \\
	&= \delta_{ij} \frac{1}{\e^{-2\beta d_i} + 1} \sum_{n_i = 0,1} \e^{-2\beta d_i n_i} n_i \\
	&= \delta_{ij} \frac{1}{\e^{-2\beta d_i} + 1} \e^{-2\beta d_i} \\
	&= \delta_{ij} \frac{1}{\e^{2\beta d_i} + 1}
\end{align}
and
\begin{align}
	\Tr{[\hat{\rho}_{\beta} \hat{g}_i \hat{g}_j^\dagger]} &= \Tr{[\hat{\rho}_{\beta} (\delta_{ij} - \hat{g}_j^\dagger \hat{g}_i)]} \\
	&= \delta_{ij} \Tr{[\hat{\rho}_\beta]} - \Tr{[\hat{\rho}_{\beta} \hat{g}_j^\dagger \hat{g}_i]} \\
	&= \delta_{ij} \left(1 -  \frac{1}{\e^{2\beta d_i} + 1} \right) \\
	&= \delta_{ij} \frac{\e^{2\beta d_i}}{\e^{2\beta d_i} + 1}
\end{align}
for all $i,j = 1,\dots,N$.
Hence, the total covariance matrix with respect to the normal modes reads
\begin{align}
	\gamma_{\beta,\hat{\bm{g}}} &= \frac{1}{\e^{2\beta D} + 1} \oplus \frac{\e^{2\beta D}}{\e^{2\beta D} + 1}.
\end{align}
In order to arrive at the covariance matrix with respect to the original modes we have to invert the normal mode transformation in \cref{Eq:fermionic_normal_mode_decomposition}, i.e., 
\begin{align}
	\gamma_{\beta, \hat{\bm{f}}} &= T^\dagger \gamma_{\beta, \hat{\bm{g}}} T.
\end{align}

Consider now the case of non-zero chemical potential.
Expanding the global number operator in the normal modes we find
\begin{align}
	\hat{N} &= \frac{1}{2} \sum_{i} (\hat{g}_i^\dagger \hat{g}_i - \hat{g}_i \hat{g}_i^\dagger).
\end{align}
and thus 
\begin{align}
	\hat{H} - \mu \hat{N} &= \sum_{i} \left(d_i - \frac{\mu}{2}\right) (\hat{g}_i^\dagger \hat{g}_i - \hat{g}_i \hat{g}_i^\dagger).
\end{align}
Hence, the results for non-zero chemical potential follow from the results for vanishing chemical potential by shifting
the normal frequencies according to $d_i \mapsto d_i - \frac{\mu}{2}$.

\subsection{Jordan-Wigner transformation}
\label{Sec:jordan_wigner_transformation}

In order to be able to deal with the fermionic ladder operators numerically, we need a matrix representation of these operators.
While for bosonic creation and annihilation operators these matrix representations are straightforward to obtain, in the fermionic case this is more involved. 
The reason for this complication is rooted in the CAR of \cref{Eq:car_elements} which ensure the antisymmetry of the fermionic wave function.
This issue can be circumvented by mapping the fermionic algebra onto the spin algebra via the Jordan-Wigner transformation~\cite{Jordan1928,Nielsen2005,Parkinson2010}.
To uniquely define the Fock space we first fix the order of the fermionic operators to be
\begin{align}
	\ket{n_1, n_2, \dots, n_N} &:= (f_1^\dagger)^{n_1}(f_2^\dagger)^{n_2} \dots (f_N^\dagger)^{n_N} \ket{\Omega}
\end{align}
where $\ket{\Omega}$ denotes the vacuum state such that $f_i\ket{\Omega} = 0$ for all $i = 1,\dots,N$.
With this order the Jordan-Wigner transformation reads
\begin{align}
	f_i &\mapsto \left( \bigotimes_{k = 1}^{i-1} \sigma_k^z \right) \otimes \sigma_i^-, & f_i^\dagger &\mapsto \left( \bigotimes_{k = 1}^{i-1} \sigma_k^z \right) \otimes \sigma_i^+
	\label{Eq:jordan_wigner_transformation_definition}
\end{align}
where
\begin{align}
	\hspace{-4pt}
	\sigma_i^+ &:= 
	\begin{pmatrix}
		0 & 0 \\
		1 & 0
	\end{pmatrix},
	\quad 
	\sigma_i^- := 
	\begin{pmatrix}
		0 & 1 \\
		0 & 0
	\end{pmatrix},
	\quad
	\sigma_i^z :=
	\begin{pmatrix}
		1 & 0 \\
		0 & -1
	\end{pmatrix}
\end{align}
are the Pauli spin operators.
Thus, after the Jordan-Wigner transformation the CAR are implicitely encoded in the algebra of the Pauli spin operators.
Equipped with transformation rules in \cref{Eq:jordan_wigner_transformation_definition} the matrix representation of beam splitters, phase shifters and two-mode squeezing operators follows straightforwardly.
Consider for example the beam splitter defined by
\begin{align}
	\hat{B}(\theta) &:= \exp{\{\ii \theta \; (\hat{f}_i \hat{f}_{i+1} - \hat{f}_i \hat{f}_{i+1})\}}.
\end{align}
Exploiting the fact that ${\sigma_k^z}^2 = \openone$ we find
\begin{align}
	\hat{f}_i^\dagger \hat{f}_{i+1} &\mapsto \left[\left( \bigotimes_{k = 1}^{i-1} \sigma_k^z \right) \otimes \sigma_i^+ \right] \left[ \left( \bigotimes_{k = 1}^{i} \sigma_k^z \right) \otimes \sigma_{i+1}^- \right] \\
	&= \left( \bigotimes_{k=1}^{i-1} {\sigma_k^z}^2 \right) \otimes \sigma_{i}^+ \sigma_{i}^{z} \otimes \sigma_{i+1}^{-} \\
	&= \sigma_{i}^{+} \otimes \sigma_{i+1}^{-}
\end{align}
and analogously
\begin{align}
	\hat{f}_i \hat{f}_{i+1}^\dagger &\mapsto - \sigma_{i}^{-} \otimes \sigma_{i+1}^+.
\end{align}
Hence, the beam splitter is mapped to the spin operator
\begin{align}
	\hat{B}(\theta) &\mapsto \exp{\{\ii \theta \; (\sigma_{i}^{+} \sigma_{i+1}^{-} + \sigma_{i}^{-} \sigma_{i+1}^{+})\}}.
\end{align}
Similarly, we obtain for the phase shifter
\begin{align}
	\hat{P}(\phi) &:= \exp{\{\ii \phi \; \hat{f}_i^\dagger \hat{f}_i\}} \mapsto \exp{\{\ii \phi \; \sigma_{i}^{+} \sigma_{i}^{-}\}}
\end{align}
and for the squeezing operator
\begin{align}
	\begin{split}
		\hat{S}(\delta) &:= \exp{\{\delta (\hat{f}_i^\dagger \hat{f}_{i+1}^\dagger - \hat{f}_i \hat{f}_{i+1})\}} \\
		&\mapsto \exp{\{\delta (\sigma_{i}^+ \sigma_{i+1}^+ + \sigma_{i}^- \sigma_{i+1}^-)\}}.	
	\end{split}
\end{align}

\section{Matrix product operators}
\subsection{Definition}

Consider a physical system comprising $N$ particles each associated with a Hilbert space of dimension $d$.
Choosing a basis $\{\ket{n_k}\}_{n_k = 1}^{d}$ of each of these Hilbert spaces a general density operator admits the form
\begin{align}
	\hat{\rho} &= \sum_{\substack{m_1, \dots, m_N = 1\\ n_1, \dots, n_N = 1}}^{d} C^{n_1,\dots,n_N}_{m_1,\dots,m_N} \ket{m_1,\dots,m_N} \bra{n_1,\dots,n_N}.
\end{align}
For fixed basis the coefficient tensor $C^{n_1, \dots, n_N}_{m_1, \dots, m_N}$ encodes all information about $\hat{\rho}$.
Unfortunately, the number of elements of this tensor scales as $d^{2N}$, i.e. exponentially in $N$.
This is known as the curse of dimensionality and constitutes a fundamental and severe challenge in classical simulations.

Matrix product operators (MPOs) aim to overcome this challenge by replacing the coefficient tensor $C^{n_1,\dots,n_N}_{m_1,\dots,m_N}$ by a product of matrices $A_{m_k}^{n_k} \in \C^{r_{k-1} \times r_k}$ such that
\begin{align}
	\hat{\rho} &= \sum_{\substack{m_1, \dots, m_N = 1\\ n_1, \dots, n_N = 1}}^{d} A_{m_1}^{n_1} A_{m_2}^{n_2} \cdots A_{m_N}^{n_N} \ket{\bm{m}} \bra{\bm{n}}.
\end{align}
where $\ket{\bm{m}} := \ket{m_1, \dots, m_N}$.
Evidently, the complexity of this alternative representation is ultimately determined by the dimensions of the matrices $A_{m_k}^{n_k}$, i.e. the values $\{r_k\}_{k=1}^{N-1}$ which are commonly referred to as bond dimensions or ranks and denoted in the main text by $D$.
Note here that we have to set $r_0 = r_N = 1$ in order to obtain a scalar from the matrix product.
Assuming that $r := \max_{k} r_k$ is bounded by a constant the number of parameters to amounts to $\LandO{r^2 d^2 N}$.
Hence, the exponential dependence on $N$ reduces to a linear dependence on $N$.

\subsection{Complexity estimates}
\label{Sec:complexity_estimates}

Computational cost is one of the key aspects in the numerical preparation of states.
In order to quantify this cost we consider the total number of floating point operations (fpos) as a figure of merit.
In the following, we derive estimates of the number of fpos for the MPO-MPO product and the SVD compression which are the main operations in imaginary time-evolution via TEBD and our proposed scheme.

\subsubsection{Dot product}

Consider two MPOs $\hat{A}$ and $\hat{B}$ of length $N$ defined by the coefficient tensors
\begin{align}
	A_{i_1,\dots,i_N}^{j_1,\dots,j_N} &:= \sum_{\alpha_1, \dots, \alpha_{N-1} = 1}^{r_1,\dots,r_{N-1}} A_{\alpha_1,i_1}^{\alpha_0, j_1} A_{\alpha_2,i_2}^{\alpha_1,j_2} \cdots A_{\alpha_{N-1},i_N}^{\alpha_{N}, j_N} \\
	B_{i_1,\dots,i_N}^{j_1,\dots,j_N} &:= \sum_{\beta_1,\dots,\beta_{N-1}}^{s_1,\dots,s_{N-1}} B_{\beta_1,i_1}^{\beta_0,j_1} B_{\beta_2,i_2}^{\beta_1,j_2} \cdots B_{\beta_{N-1},i_N}^{\beta_{N},j_N}.
\end{align}
Contraction of these two tensor over the physical legs leads to a new MPO $\hat{C}$ with coefficient tensor
\begin{align}
	C_{i_1,\dots,i_N}^{j_1,\dots,j_N} &:= \sum_{\gamma_1,\dots,\gamma_{N-1}}^{r_1 s_1,\dots,r_{N-1} s_{N-1}} C_{\gamma_1,i_1}^{\gamma_0, j_1} C_{\gamma_2,i_2}^{\gamma_1,j_2} \cdots C_{\gamma_{N-1},i_N}^{\gamma_{N},j_N}
\end{align}
where
\begin{align}
	C_{\gamma_{\ell}, i_\ell}^{\gamma_{\ell-1}, j_\ell} &= C_{(\alpha_{\ell} \beta_{\ell}), i_\ell}^{(\alpha_{\ell-1} \beta_{\ell-1}), j_\ell} := \sum_{k_\ell = 1}^{d_\ell} A_{\alpha_{\ell},i_\ell}^{\alpha_{\ell-1},k_\ell} B_{\beta_{\ell},k_\ell}^{\beta_{\ell-1},j_\ell}.
\end{align}
Hence, the number of fpos to compute $C_{\gamma_{\ell}, i_\ell}^{\gamma_{\ell-1}, j_\ell}$ counts
\begin{align}
	\text{fpos}_{\text{dot}, \ell} &= r_{\ell-1}s_{\ell-1} r_{\ell} s_{\ell} d_\ell^2 (2d_\ell - 1) \\
	&\propto r_{\ell-1}s_{\ell-1} r_{\ell} s_{\ell} d_\ell^3.
\end{align}
The total amount is found by summing over all sites, i.e.,
\begin{align}
	\text{fpos}_{\text{dot}} &= \sum_{\ell = 1}^{N} \text{fpos}_{\text{dot},\ell} \\
	&= \sum_{\ell=1}^{N} r_{\ell-1}s_{\ell-1} r_{\ell} s_{\ell} d_\ell^2 (2d_\ell - 1)
\end{align}
and thus, defining $r := \max_{i} r_i$, $s := \max_{i} s_i$ and $d := \max_{i} d_i$, the dot product is of complexity $\LandO{Nr^2 s^2 d^3}$.

\subsubsection{SVD compression}

Truncating the ranks of an MPO via SVD is in general done in two steps.
First, we have to bring the MPO into canonical form via successive QR decompositions along the chain.
This step can be omitted if the MPO is already in canonical form.
Second, each of the local tensors is compressed by a truncated SVD decomposition.

The first step involves two main operations, the QR decomposition itself,
\begin{align}
	A_{\alpha_{\ell},i_\ell}^{\alpha_{\ell-1},j_\ell} &\cong A_{(\alpha_{\ell-1} i_\ell j_\ell), \alpha_{\ell}} \\
	&= \sum_{\gamma_\ell = 1}^{r_\ell^\prime} Q_{(\alpha_{\ell-1} i_\ell j_\ell),\gamma_\ell} R_{\alpha_\ell}^{\gamma_\ell},
\end{align}
with $r_\ell^\prime := \rank{(A_{(\alpha_{\ell-1} i_\ell j_\ell), \alpha_{\ell}})}$ and the absorption of the $R$ matrix into the subsequent site
\begin{align}
	A_{\alpha_{\ell},i_\ell}^{\alpha_{\ell-1},j_\ell} &\mapsto A_{\gamma_{\ell},i_\ell}^{\alpha_{\ell-1},j_\ell} := Q_{(\alpha_{\ell-1} i_\ell j_\ell),\gamma_\ell}\\
	A_{\alpha_{\ell+1},i_{\ell+1}}^{\alpha_{\ell},j_{\ell+1}} &\mapsto A_{\alpha_{\ell+1},i_{\ell+1}}^{\gamma_{\ell},j_{\ell+1}} = \sum_{\alpha_\ell = 1}^{r_\ell} R_{\alpha_\ell}^{\gamma_\ell} A_{\alpha_{\ell+1},i_{\ell+1}}^{\alpha_{\ell},j_{\ell+1}}.
\end{align}
Given a matrix of size $m \times n$ with rank $r \leq \min{\{m,n\}}$, the QR decomposition has complexity $\LandO{m n r}$~\cite{Golub1996}.
Thus, setting $m := r_{\ell-1} d^2$, $n := r_{\ell}$ and $r := r_\ell^\prime$ yields
\begin{align}
	\text{fpos}_{\text{QR},\ell} \propto r_{\ell-1} d_\ell^2 r_{\ell} r_{\ell}^\prime.
\end{align}
Subsequently, absorbing the $R$ matrix into the next local tensor increases the fpo count by
\begin{align}
	\text{fpos}_{\text{shift QR},\ell} &= r_\ell^\prime r_{\ell+1} d_{\ell+1}^2 (2r_\ell-1) \\
	&\propto r_\ell^\prime r_{\ell+1} d_{\ell+1}^2 r_\ell. 
\end{align}

In contrast, the second step involves an additional truncation step.
After the initial SVD decomposition
\begin{align}
	A_{\alpha_{\ell},i_\ell}^{\alpha_{\ell-1},j_\ell} &\cong A_{(\alpha_{\ell-1} i_\ell j_\ell), \alpha_{\ell}} \\
	&= \sum_{\gamma_\ell = 1}^{r_{\ell}^\prime} U_{(\alpha_{\ell-1} i_\ell j_\ell),\gamma_\ell} S_{\gamma_\ell}^{\gamma_\ell} V_{\gamma_\ell, \alpha_{\ell}}^\dagger,
\end{align}
we truncate the descending singular values in $S$ at a certain index $r_{\ell}^{\prime \prime} \leq r_{\ell}^\prime$ such that 
\begin{align}
	S \in \C^{r_{\ell}^{\prime} \times r_{\ell}^{\prime}} &\mapsto \tilde{S} \in \C^{r_{\ell}^{\prime \prime} \times r_{\ell}^{\prime \prime}} \\
	U \in \C^{r_{\ell-1} d_\ell^2 \times r_{\ell}^{\prime}} &\mapsto \tilde{U} \in \C^{r_{\ell-1} d_\ell^2 \times r_{\ell}^{\prime \prime}} \\
	V \in \C^{r_{\ell}^\prime \times r_\ell^\prime} &\mapsto \tilde{V} \in \C^{r_{\ell}^\prime \times r_{\ell}^{\prime \prime}}.
\end{align}
This index $r_{\ell}^{\prime \prime}$ is either fixed or determined dynamically to achieve a certain relative error in the truncation.
In the final step we update the current and the subsequent site, i.e., 
\begin{align}
	A_{\alpha_{\ell},i_\ell}^{\alpha_{\ell-1},j_\ell} &\mapsto A_{\gamma_{\ell},i_\ell}^{\alpha_{\ell-1},j_\ell} \cong U_{(\alpha_{\ell-1} i_\ell j_\ell),\gamma_\ell}\\
	A_{\alpha_{\ell+1},i_{\ell+1}}^{\alpha_{\ell},j_{\ell+1}} &\mapsto A_{\alpha_{\ell+1},i_{\ell+1}}^{\gamma_{\ell},j_{\ell+1}} = \sum_{\gamma_\ell = 1}^{r_\ell^{\prime \prime}} \sum_{\alpha_\ell = 1}^{r_\ell^\prime} S_{\gamma_\ell}^{\gamma_\ell}V_{\gamma_\ell, \alpha_\ell}^\dagger A_{\alpha_{\ell+1},i_{\ell+1}}^{\alpha_{\ell},j_{\ell+1}}.
\end{align}
Given a matrix of size $m \times n$ the SVD decomposition has complexity $\LandO{mn \min\{m,n\}}$~\cite{Golub1996} which translates to
\begin{align}
	\text{fpos}_{\text{SVD},\ell} \propto r_{\ell-1}^\prime d_\ell^2 {r_{\ell}^\prime}^2
\end{align}
with $m := r_{\ell-1}^\prime d_\ell^2$ and $n := r_{\ell}^\prime$.
Here we exploited the fact that $\min{\{ r_{\ell-1}^\prime d_\ell^2, r_\ell^\prime \}} = r_\ell^\prime$ which is assured by the canonical form generated in the first step.
Absorbing the $S$ and $V^\dagger$ matrix into the next local tensor then involves 
\begin{align}
	\begin{split}
		\text{fpos}_{\text{shift SVD},\ell} &= r_\ell^{\prime \prime} r_\ell^{\prime} (2r_\ell^{\prime \prime}-1) \\
		&\hspace{15pt} + r_\ell^{\prime \prime} r_{\ell+1}^{\prime} d_{\ell+1}^2 (2r_\ell^{\prime}-1)
	\end{split} \\
 	&\propto {r_\ell^{\prime \prime}}^2 r_{\ell}^{\prime} +  r_\ell^{\prime \prime} r_{\ell+1}^{\prime} d_{\ell+1}^2 r_\ell^{\prime}
\end{align}
operations.
\cref{Tab:overview_fpo_counts_svd} gives an overview of the number of fpos in the individual steps of SVD compression.
The total number of fpos is finally obtained by summing over the contributions of each site.

\begin{table}[H]
\begin{center}
\caption{\textbf{Overview of fpos in SVD compression per site:} The table shows the estimated number of fpos required in the SVD compression scheme per site. Here, $r_\ell^{\prime} := \rank{(A_{(\alpha_{\ell-1} i_\ell j_\ell), \alpha_{\ell}})} \leq \min{\{r_{\ell-1} d_\ell^2, r_\ell\}}$ and $r_{\ell}^{\prime \prime}$ denotes the truncation rank.}
\label{Tab:overview_fpo_counts_svd}
\vspace{4pt}
\renewcommand{\arraystretch}{1.5}
\begin{tabular}{c | c}
	\textbf{operation} & \textbf{number of fpos} \\ \hline \hline 
	Matrix mult. & $r_\ell^\prime r_{\ell+1} d_{\ell+1}^2 r_\ell + {r_\ell^{\prime \prime}}^2 r_{\ell}^{\prime} +  r_\ell^{\prime \prime} r_{\ell+1}^{\prime} d_{\ell+1}^2 r_\ell^{\prime}$ \\ \hline
	QR & $r_{\ell-1} d_\ell^2 r_{\ell} r_{\ell}^\prime$ \\ \hline
	SVD & $r_{\ell-1}^\prime d_\ell^2 {r_{\ell}^\prime}^2$
\end{tabular}
\end{center}
\end{table}

\section{Implementation of Bogoliubov transformations as optical circuits}
\label{Sec:implementation_bogoliubov_transformation}

The Bloch-Messiah transformation enables the construction of optical circuits that can be implemented efficiently as tensor-network transformations. 
Here we detail the optical circuit returned by the Bloch-Messiah transformation and the procedure used to obtain the parameters of this circuit.

For the bosonic setting, the Bloch-Messiah transformation decomposes a Gaussian unitary transformation into a sequence of three transformations, firstly a passive linear transformation, secondly a tensor product of single-mode squeezing transformations acting on each mode, and finally, another passive linear transformation.
The fermionic setting is similar except that the single-mode squeezing is replaced by either two-mode squeezing or a swap gate.

The action of the squeezing transformations on an MPO is straightforward to compute as these are tensor products of transformations, each of which acts on one or two modes.
The passive optical circuits too can be decomposed into a sequence of nearest-neighbour optical transformation.
The procedures for obtaining these transformations are well known in quantum-optics literature~\cite{Reck1994a,Clements2016a,Guise2018,Dhand2015a,Su2019a,Kumar2020}.
While the procedures of Refs.~\cite{Reck1994a,Clements2016a,Guise2018} enable decomposing any given $n$-mode unitary transformations into two-mode transformations, those of Refs.~\cite{Dhand2015a,Su2019a,Kumar2020} enable decomposition into $m$-mode transformations for $m<n$.

\begin{figure}
	\subfloat[]{\includegraphics[scale=0.7]{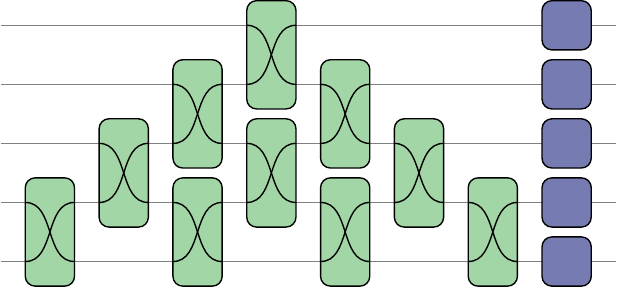} \label{Fig:reck_decomposition}}
	\hfill
	\subfloat[]{\includegraphics[scale=0.7]{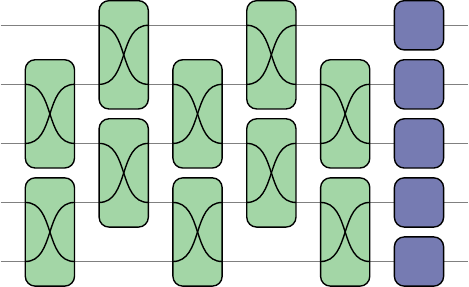} \label{Fig:clements_decomposition}}
	\caption{Decomposition of passive linear optical transformation into beam-splitter and phase-shifter transformations via the triangular scheme of Reck \textit{et al.} (left) and via the rectangular procedure of Clements \textit{et al.} (right).
	The green boxes represent two-mode beam-splitter transformations, while the blue dots represent single-mode phase-shifter transformations}
	\label{Fig:decomposition_schemes}
\end{figure}

For concreteness, we focus on decompositions that we exploited in the simulations detailed in the main text, i.e., decomposition into two-mode (beam-splitter) and single-mode (phase-shifter) transformations.
The two-mode decompositions receive as input an $n\times n$ special unitary matrix, which describes the action of the passive linear transformation on $n$ bosonic modes.
The procedures returns a sequence of two-mode beam-splitter transformations and phase-shifter transformations.
To obtain these transformations, each element of the given unitary matrix is \textit{nulled} systematically in a manner similar to Gaussian elimination to convert the given unitary matrix into a diagonal unitary matrix.
The circuits obtained from the Reck~\textit{et al.} and the Clements~\textit{et al.} decompositions are depicted in~\cref{Fig:decomposition_schemes}.
The circuits for the de Guise \textit{et al.} decomposition are similar to those of the Reck \textit{et al.} in the location of the beam-splitter transformation but differs in where the phase-shifters are placed.
A detailed description of the procedure to obtain the circuit parameters is provided in the Appendix of Ref.~\cite{Kumar2020}, and code to obtain these parameters is available online~\cite{Killoran2019a}.

In our simulations, we explored the Reck \textit{et al.}~and the Clements \textit{et al.} decompositions.
I.e., we either used the Reck \textit{et al.}~procedure for both the passive linear unitary transformations or the Clements \textit{et al.}~decomposition for both.
The triangular structure of the Reck \textit{et al.}~decomposition led to lower intermediate bond dimensions than the Clements \textit{et al.}~decomposition, which possesses a rectangular structure. 
This leads to significantly lower computational cost in using the Reck \textit{et al.}~decomposition.
We note that in using the Reck \textit{et al.}~decomposition, we arrange the normal modes such that the most highly populated modes (i.e., the ones with the lowest frequency) are incident on the shallowest portion of the circuit with depth unity while the least populated modes are incident on the deepest part of the circuit with depth $2N-3$.
This point is detailed in the last paragraph of~\cref{Sec:supplementary_spin_boson_model}.
One potential direction to explore is whether using different decompositions for the two passive linear unitary transformations leads to improved scaling as compared to using the same decomposition for both.

\section{Time-Evolving Block Decimation}
\label{Sec:tebd}

In this section we briefly introduce the Time-Evolving Block Decimation (TEBD) algorithm which serves as a performance benchmark for our method.
For a more detailed treatment we refer to Refs.~\cite{Vidal2003,Vidal2004,Paeckel2019}.
TEBD is an evolution scheme that is specifically tailored to Hamiltonians whose interaction terms have a local structure. 
Consider therefore an $N$-particle Hamiltonian of the form
\begin{align}
	\hat{H} &:= \sum_{k = 0}^{N-1} \hat{h}_{k, k+1}
\end{align}
where $\hat{h}_{k, k+1}$ is a local Hamiltonian acting only on site $k$ and $k+1$.
Note that both Hamiltonians presented in the main text admit this form with a proper choice of the two-site terms $\hat{h}_{k, k+1}$.

TEBD provides a way to implement propagators of the form $\e^{-\beta \hat{H}}$ as tensor-networks based on the Suzuki-Trotter decomposition of operator exponentials.
In particular, we proceed in three steps.
Firstly, we decompose the total propagation into $n$ small steps of size $\delta \beta := \frac{\beta}{n} \ll 1$ such that 
\begin{align}
	\hat{U}(\beta) &:= \e^{-\beta \hat{H}} = \left(\e^{-\delta \beta \hat{H}}\right)^{n}.
\end{align}
Secondly, for reasons that become clear in the following, we split the sum defining the global Hamiltonian into an even and an odd part such that
\begin{align}
	\hat{H} &= \sum_{k \text{ even}} \hat{h}_{k, k+1} + \sum_{k \text{ odd}} \hat{h}_{k, k+1} =: \hat{H}_{\text{even}} + \hat{H}_{\text{odd}}.
\end{align}
Finally, based on this splitting of the Hamiltonian we apply a Suzuki-Trotter decomposition to the propagator $\hat{U}(\delta \beta) := \e^{-\delta \beta \hat{H}} = \e^{- \delta \beta (\hat{H}_{\text{even}} + \hat{H}_{\text{odd}})}$.
In principle, this decomposition can be carried out up to an arbitrary order in $\delta\beta$~\cite{Dhand2014}.
However, in this work we constrain ourselves to the most common choices which are first, second and fourth order for reasons of numerical stability.
The first order Suzuki-Trotter decomposition takes the simple form
\begin{align}
	\hat{U}_1(\delta\beta) &:= \e^{-\delta\beta \hat{H}_{\text{even}}} \, \e^{-\delta\beta \hat{H}_{\text{odd}}}
\end{align}
such that $\hat{U}(\delta\beta) = \hat{U}_1(\delta\beta) + \LandO{\delta\beta^2}$ for $\delta\beta \rightarrow 0$.
A second order Suzuki-Trotter decomposition can be achieved by combining a first order propagator with its reversed counterpart leading to
\begin{align}
	\hat{U}_2(\delta\beta) &:= \e^{-\frac{\delta\beta}{2} \hat{H}_{\text{even}}} \, \e^{-\frac{\delta\beta}{2} \hat{H}_{\text{odd}}} \e^{-\frac{\delta\beta}{2} \hat{H}_{\text{odd}}} \, \e^{-\frac{\delta\beta}{2} \hat{H}_{\text{even}}} \\
	&=  \e^{-\frac{\delta\beta}{2} \hat{H}_{\text{even}}} \, \e^{-\delta\beta \hat{H}_{\text{odd}}} \e^{-\frac{\delta\beta}{2} \hat{H}_{\text{even}}}.
\end{align}
As the name already suggests the second order Suzuki-Trotter decomposition fulfills $\hat{U}(\delta\beta) = \hat{U}_2(\delta\beta) + \LandO{\delta\beta^3}$ for $\delta\beta \rightarrow 0$.
Finally, the fourth order Suzuki-Trotter decomposition is constructed by combining five second order propagators. 
More precisely, defining
\begin{align}
	\delta\beta_1 &:= \frac{1}{4 - 4^{\frac{1}{3}}} \; \delta\beta \\
	\delta\beta_2 &:= \delta\beta - 4 \delta\beta_1
\end{align}
we obtain
\begin{equation}
	\hat{U}_{4}(\delta\beta) := \hat{U}_2(\delta\beta_1) \hat{U}_2(\delta\beta_1) \hat{U}_2(\delta\beta_2) \hat{U}_2(\delta\beta_1) \hat{U}_2(\delta\beta_1)
\end{equation}
such that $\hat{U}(\delta\beta) = \hat{U}_4(\delta\beta) + \LandO{\delta\beta^5}$ for $\delta\beta \rightarrow 0$.

The previous considerations suggest that the error made by replacing $\hat{U}(\beta)$ by a sequence $(\hat{U}_{k}(\delta \beta))^{n}$ with $k = 1,2,4$ solely depends on $\delta \beta^{k+1}$.
In fact, it additionally depends on the Hamiltonian and its splitting respectively.
More precisely, it has been shown recently~\cite{Childs2020} that there exists $C > 0$ and $\delta\beta_0 > 0$ such that
\begin{align}
	\|\hat{U}(\delta \beta) - \hat{U}_{k}(\delta \beta)\|_2 \leq C \; \alpha \; \delta\beta^{k+1} \; \e^{4\delta\beta \Upsilon (\|\hat{H}_{\text{even}}\|_2 + \|\hat{H}_{\text{odd}}\|_2)}
	\label{Eq:suzkuki_trotter_error}
\end{align}
for all $\delta \beta < \delta \beta_0$.
Here, we have defined
\begin{align}
	\alpha &:= \sum_{\substack{\gamma_1, \dots, \gamma_{k+1} \\ = \text{ even},\text{ odd}}} \left\|\left[\hat{H}_{\gamma_{k+1}}, \dots [\hat{H}_{\gamma_2}, \hat{H}_{\gamma_1}]\right]\right\|_2  , \\
 	\Upsilon &:= 
	\begin{cases}
		1  							& \text{for } k = 1 \\
		2 \cdot 5^{\frac{k}{2}-1}   & \text{for } k > 1,
	\end{cases}
\end{align}
and $\| \cdot \|_2$ denotes the spectral norm.
Hence, in general, it is highly non-trivial to choose the best Suzuki-Trotter decomposition for a given task.
However, \cref{Eq:suzkuki_trotter_error} can be used to deduce two rules of thumb.
Firstly, as the inverse temperature $\beta$ increases, a higher number of Trotter steps $n$ is required to guarantee a specific Trotter error.
Secondly, as the number of constituent $N$ increases, in most physical systems also the spectral norms $\| \hat{H}_{\text{even}} \|_2$ and $\| \hat{H}_{\text{odd}}\|_2$ as well as the commutator norms in $\alpha$ increase.
Thereby, more Trotter steps $n$ are necessary to meet a certain error threshold.

\begin{figure}[H]
	\includegraphics[width=1.0\linewidth]{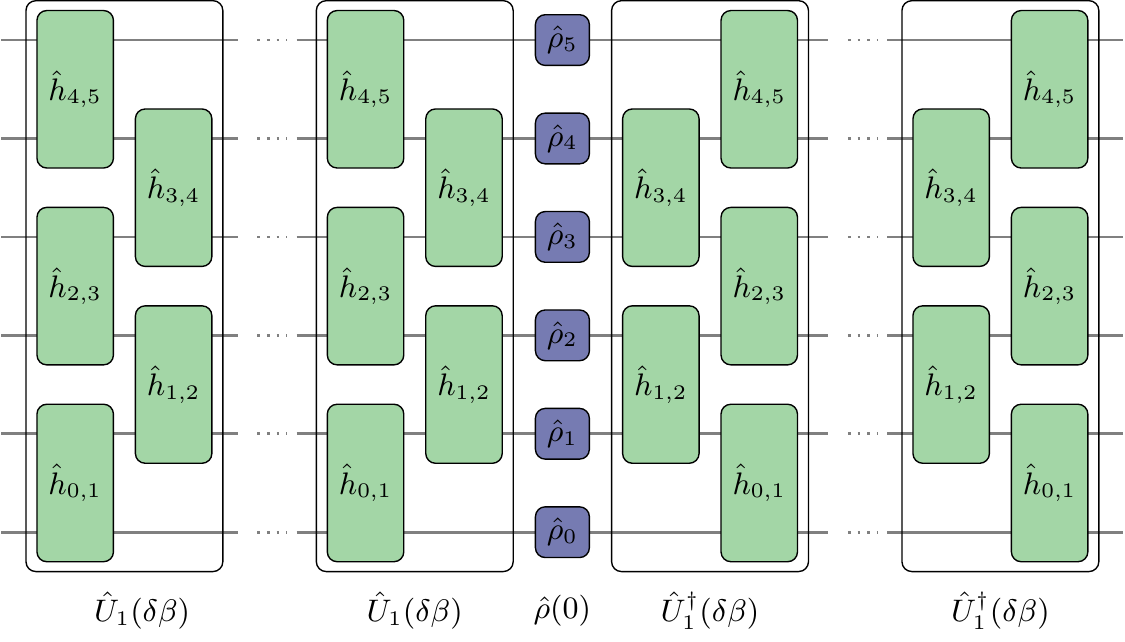}
	\caption{\textbf{Tensor-network representation of the evolution of a density operator via TEBD in imaginary time:}
	Schematic representation of the tensor-network structure arising from evolving a product state $\hat{\rho}(0)$ via TEBD in imaginary time $\beta$.
	The final state is obtained by successively applying the first order propagator $\hat{U}_1(\delta\beta)$ to $\hat{\rho}$ from the left and the right.
	Here, each $\hat{U}_1(\delta\beta)$ consists of two layers comprising only two-site gates.
	For higher order propagators the number of layers within one time step increases.
	}
	\label{Fig:tebd_schematic}
\end{figure}

Once a specific decomposition is chosen the evolution of the state follows by successive application of this propagator to the density operator from both sides.
Since terms within $\hat{H}_{\text{even}}$ commute amongst each other we can expand exponentials of the form $\e^{-\delta\beta \hat{H}_{\text{even}}}$ into
\begin{align}
	\e^{-\delta\beta \hat{H}_{\text{even}}} &:= \prod_{k \text{ even}} \e^{-\delta\beta \hat{h}_{k, k+1}}
	\label{Eq:even_propagator_product}
\end{align}
without introducing any additional Suzuki-Trotter error.
Thus, the global operator $\e^{-\beta \hat{H}_{\text{even}}}$ is just a tensor product of the local two-site gates $\e^{-\beta \hat{h}_{k, k+1}}$.
In fact, this is the reason to separate even and odd parts of the Hamiltonian in the first place.
Naturally, the same reasoning also applies to $\hat{H}_{\text{odd}}$.
Hence, assuming a finite Hilbert space dimensions $M$ for each particle, \cref{Eq:even_propagator_product} and its counterpart can be represented as an MPO with bond dimension at most $r = M^2$.
The tensor network arising from applying the sequence of propagators to the density operators is exemplified for the first order Suzuki-Trotter decomposition in \cref{Fig:tebd_schematic}.
From an implementation point of view it is important to note that we compress the state after each MPO-MPO product instead of compressing only after a fully completed time step. 

\section{Numerical examples}
\label{Sec:numerical_examples}

In this section we provide all the details concerning the numerical examples in the main text.

\subsection{Figures of merit}
\label{Sec:figures_of_merit}
As we pointed in the main text, when preparing a state numerically we have two competing figures of merit. 
On the one hand we want to prepare the state with a specific accuracy and on the other hand we want to do this with the least computational effort.

Consider first the issue of measuring the accuracy.
In this work we focus on the preparation of Gaussian states which are determined by the first and second moments of the ladder operators.
As a consequence, very natural distance measures such as the fidelity~\cite{Banchi2015} can also be expressed as a function of these two moments.
This suggests to use the fidelity of the exact and the prepared state to measure the faithfulness of the respective preparation scheme.
However, neither imaginary time evolution nor our scheme ensures positivity of the first and second moments which prevents the use of this Gaussian fidelity formula.
Hence, we measure the accuracy of the prepared state by the norm difference in first and second moments.
More precisely, as we are primarily interested in undisplaced Gaussian states which have vanishing first moments we consider the absolute Frobenius norm, $\epsilon_{\bm{m}} := \|\bm{m}_{\beta, \hat{\bm{a}}}\|_{\text{F}}$, for the first moments and the relative Frobnius norm difference
\begin{align}
	\epsilon_{\gamma}^{\text{rel}} &:= \frac{\| \gamma_{\beta,\hat{\bm{a}}} - \gamma^{\text{ex}}_{\beta,\hat{\bm{a}}} \|_{\text{F}}}{\|\gamma^{\text{ex}}_{\beta,\hat{\bm{a}}}\|_{\text{F}}}
\end{align}
for the second moments.
In all examples provided in this work we demand these errors to be below 1\%, i.e., $\epsilon_{\bm{m}} < 10^{-2}$ and $\epsilon_{\gamma}^{\text{rel}} < 10^{-2}$. 
Strictly speaking, this is a flawed measure of accuracy since again neither imaginary time evolution nor our preparation scheme ensures strict Gaussianity of the final state.
Thus, one would have to check all higher moments of the state as well which is of course impossible.
Nevertheless, we want to emphasize that the non-Gaussianity of the final state is merely a by-product of the imprecision of the truncated initial states and propagators.
Hence, since we are able to systematically increase the precision of the initial states as well as of the evolutions, we are also able to control the Gaussianity of the final state.

In order to quantify the computational effort, we estimate the number of floating point operations (fpos) rather than measuring the CPU time.
Counting the number of fpos required to perform a specific task has the advantage of being independent of both the specific implementation and the underlying hardware.
From our point of view, this allows us to get a much clearer idea of the real complexity of the considered algorithm.
In our application we essentially have to keep track of two operations, the product of two MPOs and the compression of an MPO.
While there are many ways to compress an MPO~\cite{Schollwock2011a}, we stick to compression via standard SVD in this work.
The complexity estimates for these two operations are derived in detail in \cref{Sec:complexity_estimates}.
Equipped with these estimates the total fpo count of the state preparation is then obtained by logging the number of fpos in each step of the algorithm.

\subsection{Spin-boson model with Ohmic spectral density}
\label{Sec:supplementary_spin_boson_model}

\begin{figure}[H]
	\includegraphics[width=\linewidth]{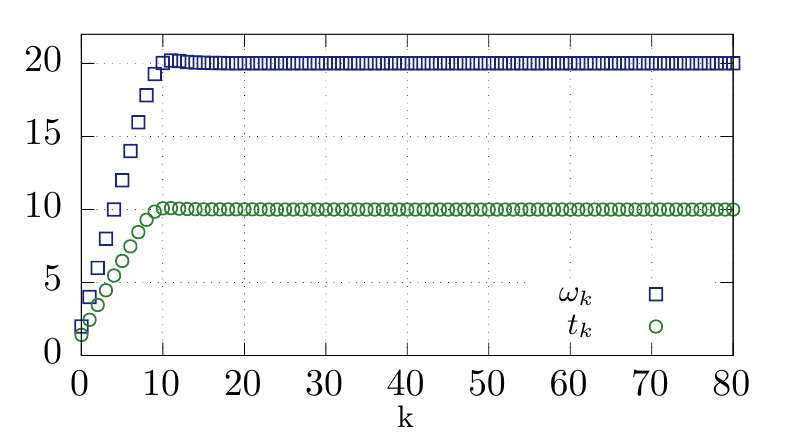}
	\caption{\textbf{Discrete frequencies and couplings of discretized spin-boson model:}
	Discrete frequencies and couplings obtained after TEDOPA chain mapping.
	These frequencies are computed using the routine described in Ref.~\cite{Gautschi1994}.
	}
	\label{Fig:TEDOPA_freqs_and_couplings}
\end{figure}

As a bosonic example we chose the spin-boson model with an Ohmic spectral density.
This model is considered in a wide variety of applications~\cite{Weiss2012,Leggett1987} and aims to describe a two-level system coupled to a bosonic environment.
In particular, the Hamiltonian of this model splits into three terms $H := H_{\text{sys}} + H_{\text{env}} + H_{\text{int}}$.
Here, $H_{\text{sys}}$ denotes the arbitrary free system Hamiltonian and $H_{\text{env}}$ the free environment Hamiltonian defined by 
\begin{align}
	\hat{H}_{\text{env}} &:= \int_{\omega_{\min}}^{\omega_{\max}} \mathrm{d} \omega \;\omega\, \hat{a}_\omega^\dagger \hat{a}.
\end{align}
The coupling of the system to the environment is governed by the interaction Hamiltonian
\begin{align}
	\hat{H}_{\text{int}} &:= \hat{A}_{\text{sys}} \; \int_{\omega_{\min}}^{\omega_{\max}} \mathrm{d}\omega \; \sqrt{\frac{J(\omega)}{\pi}} (\hat{a}_\omega^\dagger + \hat{a}_\omega)
\end{align}
where $A_{\text{sys}}$ denotes an arbitrary coupling operator of the system and 
\begin{align}
	J: [\omega_{\min}, \omega_{\max}] \rightarrow [0, \infty),\; \omega \mapsto \omega \; \e^{-\omega}
\end{align}
denotes the Ohmic spectral density.
Furthermore, in all our examples we chose $\omega_{\min} = 0$ and $\omega_{\max} = 40$ which guarantees $J(\omega_{\max}) < 10^{-14}$.
In applications one is commonly interested in the dynamics of the system starting from an initial product state of the form $\hat{\rho}_0 := \hat{\rho}_{\text{sys},0} \otimes \hat{\rho}_{\text{env},0}$ where $\hat{\rho}_{\text{env},0}$ is either the ground state or a thermal state of the free environment Hamiltonian.
For simulation purposes, the continuous Hamiltonian is usually mapped to a discrete Hamiltonian that captures the relevant features.
One way of doing this is to apply the TEDOPA chain mapping~\cite{Prior2010,Chin2010} which ultimately leads to a unitarily equivalent and discrete chain Hamiltonian of the form $\hat{H}^\prime := \hat{H}_{\text{sys}} + \hat{H}_{\text{env}}^\prime + \hat{H}_{\text{int}}^\prime$ where
\begin{align}
	\hat{H}_{\text{env}}^\prime &:= \sum_{k = 0}^{\infty} \omega_k \hat{a}_{k}^\dagger \hat{a}_k + \sum_{k = 0}^{\infty} t_{k+1} (\hat{a}_{k+1}^\dagger \hat{a}_{k} + \text{h.c.})
	\label{Eq:semi_infinite_chain_hamiltonian}
\end{align}
and
\begin{align}
	\hat{H}_{\text{int}} &:= t_0 A_{\text{sys}} \; (\hat{a}_0^\dagger + \hat{a}_0).
\end{align}
In order to obtain a finite Hamiltonian we truncate the chain in \cref{Eq:semi_infinite_chain_hamiltonian} at a certain length $N$ which then results in~\cref{Eq:spin_boson_bath_hamiltonian}.
Here, the frequencies $\{\omega_k\}$ and couplings $\{t_k\}$ are determined by the three-term recurrence coefficients of the set of polynomials orthogonal with respect to the scalar product induced by the spectral density.
At this point we refer to Refs.~\cite{Prior2010,Chin2010} for a more comprehensive introduction.
In practice, these coefficients can be obtained using numerically stable routines \cite{Gautschi1994}.
We depict the first few frequencies and couplings for our specific spectral density in \cref{Fig:TEDOPA_freqs_and_couplings}.
In the example we now aim to prepare the initial thermal states of the finite and discrete bath Hamiltonian in~\cref{Eq:spin_boson_bath_hamiltonian}.

As we pointed out in the main text this requires the truncation of the infinite-dimensional Hilbert space to a finite size $M$.
In principle, this $M$ could be arbitrarily low.
However, given a certain error threshold $\epsilon$ we are able to determine reasonable lower bounds on $M$ as a function of the inverse temperature.
The underlying rationale is as follows.
Assuming a perfect implementation of the normal mode transformation $T$ the truncation has to be at least such that the first and second moments of the truncated oscillators meet the accuracy constraint, i.e.,
\begin{align}
	\| \bm{m}_{\hat{\bm{a}}}^{\text{trunc}} \|_{\text{F}} &< \epsilon &&\text{and} & \frac{\| \gamma_{\beta, \hat{\bm{a}}}^{\text{trunc}} - \gamma_{\beta,\hat{\bm{a}}}^{\text{ex}}\|_{\text{F}}}{\|\gamma_{\beta, \hat{\bm{a}}}^{\text{ex}}\|_{\text{F}}} &< \epsilon.
	\label{Eq:constraints_local_dimension}
\end{align}
In order to find a solution to this set of non-linear inequalities we start by replacing the normal mode density operator in~\cref{Eq:bosonic_normal_mode_thermal_state_fock_basis} by
\begin{align}
	\hat{\rho}_{\beta, \bm{\hat{b}}}^{\text{trunc}} &:= \frac{1}{\mathcal{Z}^{\text{trunc}}} \left[\bigotimes_{i = 1}^{N} \sum_{n_i = 0}^{M-1} \e^{-2\beta d_i n_i} \ket{n_i}_{\bm{\hat{b}}} \bra{n_i}_{\bm{\hat{b}}} \right]
\end{align}
with $\mathcal{Z}^{\text{trunc}}$ such that $\Tr{\rho_{\beta,\hat{\bm{b}}}^{\text{trunc}}} = 1$.
Equipped with this truncated density operator the first and second moments $\bm{m}_{\hat{\bm{b}}}^{\text{trunc}}$ and $\gamma_{\beta, \hat{\bm{b}}}^{\text{trunc}}$ with respect to the normal mode basis are straightforwardly obtained following the derivation in~\cref{Sec:bosonic_thermal_states}.
Applying the inverse normal mode transformation $T^{-1}$ finally leads to the expressions for $\bm{m}_{\hat{\bm{a}}}^{\text{trunc}}$ and $\gamma_{\beta, \hat{\bm{a}}}^{\text{trunc}}$.
For fixed temperature $\beta$ this procedure can be performed numerically for increasing values of $M$ until the constraints in~\cref{Eq:constraints_local_dimension} are met.
The lower bounds on $M$ obtained for our specific example are summarized in \cref{Tab:lower_bounds_local_dims}.
In practice, however, a higher value of $M$ might be required in intermediate steps of the evolution. 

\begin{table}[H]
	\begin{center}
	\caption{\textbf{Lower bound on local dimension as a function of the rescaled inverse temperature:}
	Denoting the untruncated and truncated covariance matrix by $\gamma_{\beta, \bm{\hat{b}}}^{\text{ex}}$ and $\gamma_{\beta, \bm{\hat{b}}}^{\text{trunc}}$, respectively, the table depicts the minimal local Hilbert space dimension required such that $\|\bm{m}_{\beta,\bm{\hat{b}}}^{\text{trunc}} \| < 10^{-2}$ as well as $\|\gamma_{\beta, \bm{\hat{b}}}^{\text{trunc}} - \gamma_{\beta, \bm{\hat{b}}}^{\text{ex}}\| \|\gamma_{\beta, \bm{\hat{b}}}^{\text{ex}}\|^{-1} < 10^{-2}$.}
	\label{Tab:lower_bounds_local_dims}
	\setlength{\tabcolsep}{10pt}
	\renewcommand{\arraystretch}{2.0}
		\begin{tabular}{c || c | c | c | c | c}
			$\beta_{\text{resc}}$ & 0.5 & 1.0 & 2.0 & 3.0 & 4.0 \\\hline			
			$M$ & 14 & 8 & 4 & 3 & 3
		\end{tabular}
	\end{center}
\end{table}

Based on these preliminaries the results depicted in \cref{Fig:fpos_vs_beta} and \cref{Fig:modes_vs_fpos} are obtained by optimizing over the various simulation parameters such that $\epsilon_{\bm{m}} < 10^{-2}$ and $\epsilon_{\gamma}^{\text{rel}} < 10^{-2}$ holds and the number of fpos is minimal.
Here, the different preparation schemes involve different number of optimization parameters.
The ones that both have in common are the local Hilbert space dimension as well as the compression parameters, i.e., relative truncation error $\epsilon_{\text{rel}}$ and maximal truncation rank $r_{\text{max}}$, respectively.
While we explored different state compression settings in both schemes, we found that the single gates of the Gaussian circuit are hardly compressible and thus omitted gate compression here.
For imaginary time evolution, however, we also compressed the single gates with a relative truncation error of $\epsilon^{\text{gate}}_{\text{rel}} = 10^{-7}$.
In the Gaussian scheme we have only one additional parameter, the type of circuit decomposition, whereas for imaginary time evolution we have two additional parameters, namely the order of the Suzuki-Trotter decomposition and the time step size.
Finally, we want to emphasize that we do not consider a solution to be optimal if it meets the accuracy constraint but is not converged.
That means that an optimal solution should always be such that the accuracy increases when the approximations are relaxed.
This excludes undesired solutions that arise from cancellation of different approximation errors.

\begin{figure}[H]
	\begin{center}
		\includegraphics[width=\linewidth]{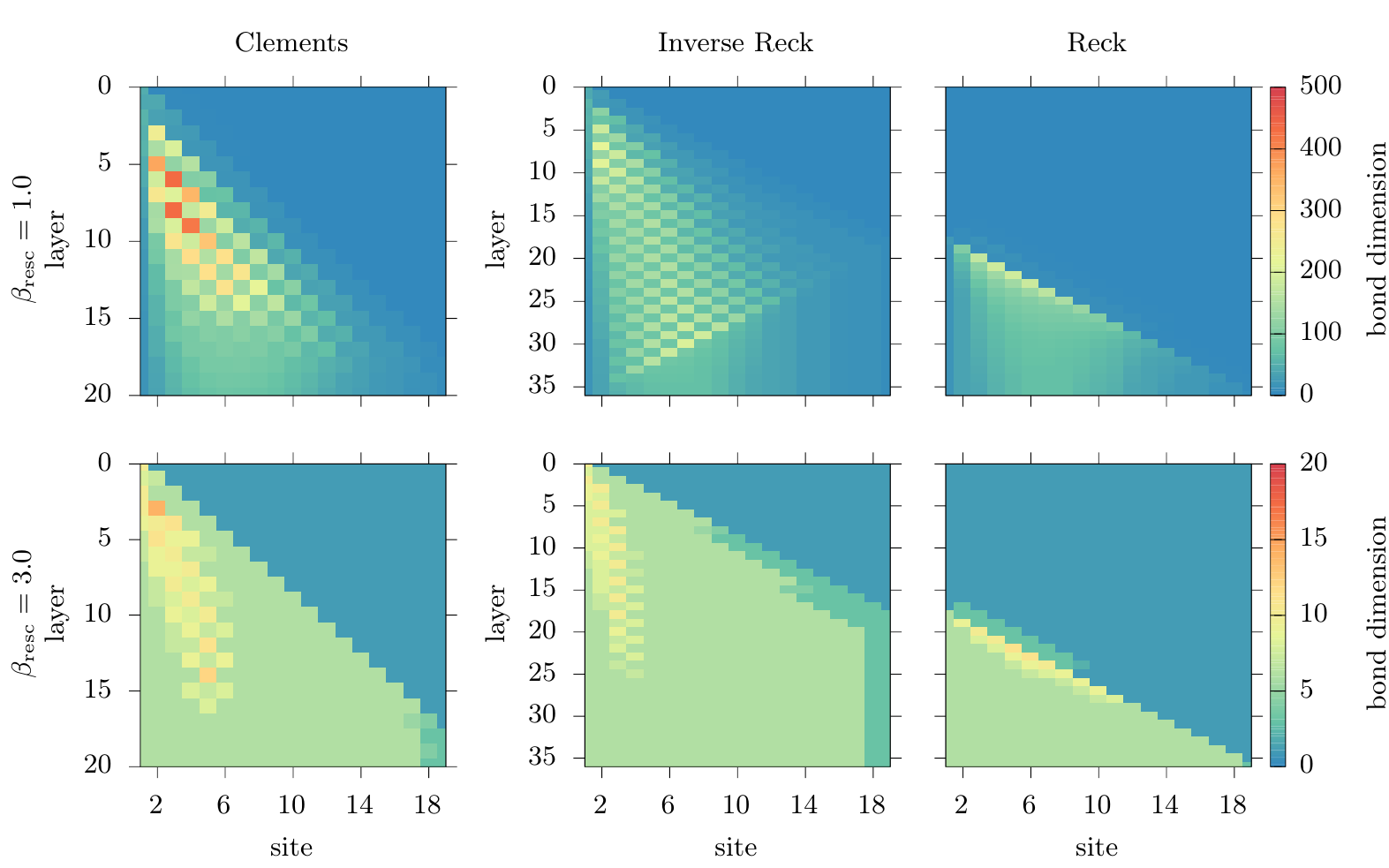}
		\caption{\textbf{Evolution of bond dimension in spin-boson model for different circuit decompositions:}
		Comparison of the evolution of bond dimensions in the preparation of a thermal state of the spin-boson model for the Clements, the Reck and the inverse Reck circuit decomposition.
		Here, the initial state is constructed such that average occupation per site decreases with increasing site index.
		For both temperatures, $\beta_{\text{resc}} = 1.0$ and $\beta_{\text{resc}} = 3.0$, the system comprises $N = 20$ modes and we fix a relative truncation error $\epsilon^{\text{state}}_{\text{trunc}} = 10^{-5}$, a cut-off frequency $\omega_c = 1$ and a hard cut-off at $\lambda \omega_c = 40$.
		The local dimension is chosen as $M = 8$ and $M = 4$, respectively. 
		Note that we did not constrain the maximal bond $r^{\text{state}}_{\text{max}}$ here.
		}
		\label{Fig:heatmap_ranks_different_circuits}		
	\end{center}
\end{figure}

The first question we consider is how the number of fpos scales as function of the rescaled inverse temperature $\beta_{\text{resc}}$.
\cref{Tab:sbm_overview_beta_vs_fpos_gauco} and \cref{Tab:sbm_overview_beta_vs_fpos_itime} summarize the optimization parameters for the Gaussian preparation and imaginary time evolution respectively.
Here, we highlight the best configuration for each $\beta_{\text{resc}}$ ultimately leading to \cref{Fig:fpos_vs_beta}.
It shows that the Gaussian scheme exhibits a drastically improved performance compared to imaginary time evolution over a wide range of temperatures.
Furthermore, the improvement becomes even more dominant in the low temperature regime.
This can be explained in two ways. 
On the one hand, the larger $\beta_{\text{resc}}$ the longer we have to evolve the state in imaginary time evolution.
In order to keep the Trotter error on a reasonable level the number of Trotter steps has to increase accordingly.
Hence, even though the required Hilbert space dimension as well as the amount of correlations drops, the computational cost remains immense.
On the other hand, as $\beta_{\text{resc}}$ increases the population in the initial state of the Gaussian scheme decreases, i.e. less excitations get injected into the circuit.
As a consequence we observe less correlations building up in the circuit during preparation.
This effect is further illustrated in~\cref{Fig:heatmap_ranks_different_circuits}. 

\begin{figure}[H]
	\begin{center}
		\includegraphics[width=\linewidth]{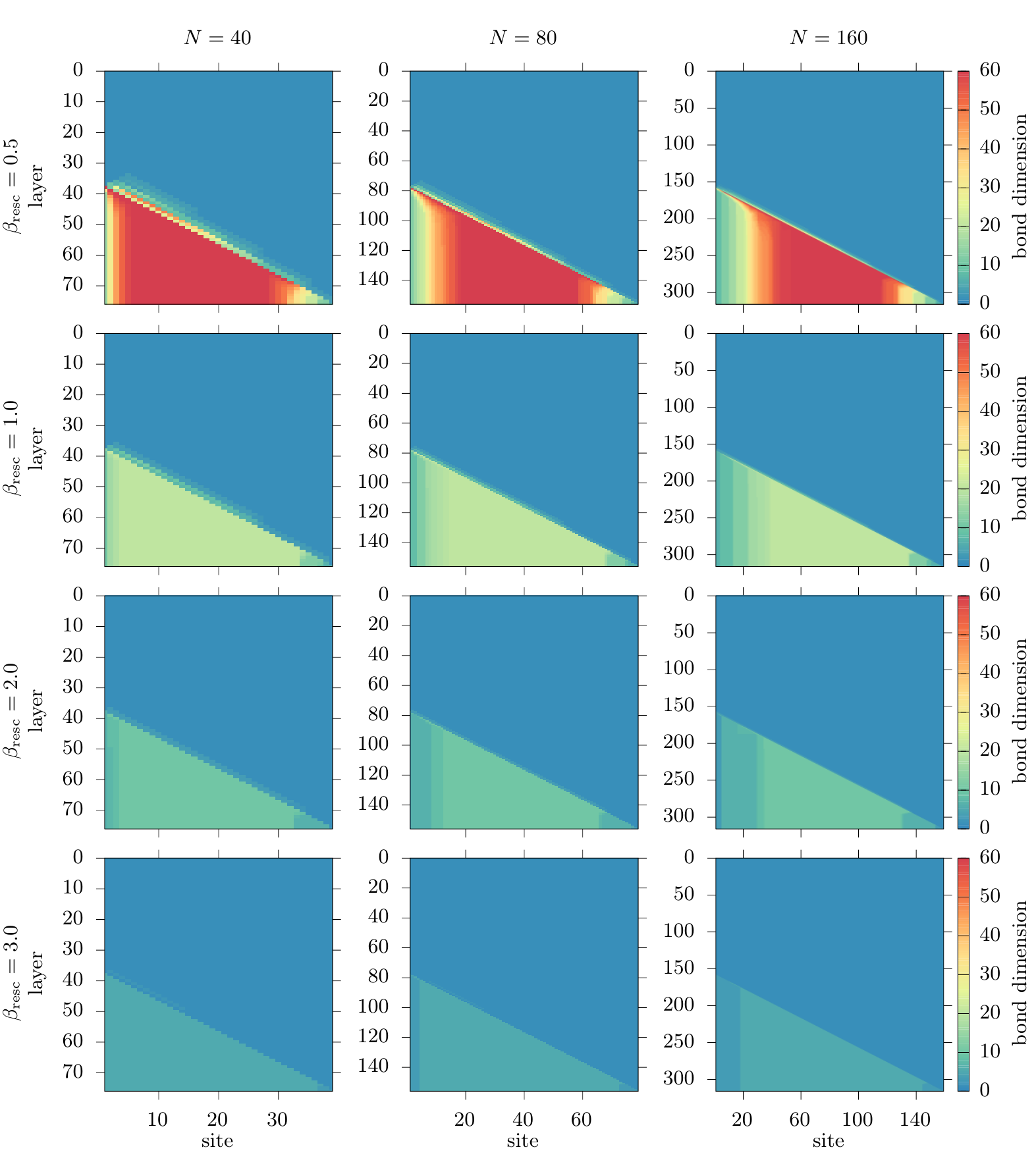}
		\caption{\textbf{Evolution of the bond dimensions in the spin-boson model for an increasing number of modes:}
		Comparison of the evolution of bond dimensions in the preparation of a thermal state of the spin-boson model for different number of modes using the Reck decomposition.
		For each $\beta_{\text{resc}}$ we depict the bond dimensions of the best configuration highlighted in \cref{Tab:sbm_overview_modes_vs_fpos_gauco}.
		}
		\label{Fig:heatmap_ranks_different_number_modes}		
	\end{center}
\end{figure}

The second question we address is how the computational cost scales with the number of modes $N$ for fixed temperature $\beta_{\text{resc}}$.
We summarize the considered optimization parameters in \cref{Tab:sbm_overview_modes_vs_fpos_gauco} and again highlight the best configuration leading to \cref{Fig:modes_vs_fpos} in boldface.
As expected the number of fpos scales polynomially in $N$ with polynomial degree $\alpha < 2$.
Interestingly, we observe that $\alpha$ decreases as $\beta_{\text{resc}}$ increases, i.e. as the temperature increases the scaling becomes more favorable.
Providing a rigorous explanation for this behavior is challenging.
However, the evolution of the bond dimensions for different temperature and different number of modes depicted in \cref{Fig:heatmap_ranks_different_number_modes} gives an intuitive idea of what is happening.
For very low temperatures the bond dimensions are almost constant along the chain and throughout all layers of the circuit irrespective of the number of modes.
This has two implications.
Firstly, each layer contributes roughly the same amount of fpos.
Secondly, for each layer the commonly used complexity estimate $\LandO{N d^2 r^3}$ with $r := \max_{\ell} r_\ell$ actually becomes a reasonably tight upper bound.
Hence, when doubling the number of modes $N$, not only the number of layers but also the computational cost in each layer doubles.
Thus, the overall number of fpos is expected to square.
For high temperatures the situation is different. 
Here, the bond dimension are no longer comparable neither across the different sites nor throughout the different layers.
Hence, the above reasoning breaks down and it is hard to deduce any kind of prediction about the scaling in $N$.

Finally, we comment on the optimal choice of the circuit decomposition. 
\cref{Tab:sbm_overview_beta_vs_fpos_gauco} indicates that in the considered range of temperatures the Reck circuit is more efficient than the Clements circuit.
At first sight this seems to be surprising since for fixed number of modes $N$ the number of layers in the Reck circuit is $2N-3$ whereas for the Clements circuit it is $N$.
Hence, one might naively expect that fewer layers also means less fpos.
However, as \cref{Fig:heatmap_ranks_different_circuits} shows the amount of correlations building up in the circuit can be tremendously different.
Here, we looked at the following situation.
We prepared the initial product state such that the average occupation of each site decreases with increasing site index.
Subsequently, we evolved this state under three different circuits, namely the Clements, the Reck and the inverse Reck circuit.
The Reck and the inverse Reck circuit differ in the orientation of the triangle of beam splitters with respect to the distribution of the initial occupations.
While for the standard Reck decomposition the depth of the circuit increases with increasing site index, for the inverse Reck decomposition it is vice versa.
As a consequence, highly populated modes propagate through a lot more beam splitters and phase shifters in the inverse Reck circuit than in the standard Reck circuit.
This leads to a lot more correlations in the early stages of the preparation scheme which are then spread across the modes.
Thus, in this specific scenario the standard Reck circuit proves to be more favorable than the other circuits.
Solving the question of the optimal order of the initial state as well as the optimal circuit decomposition in a general scenario is deferred to future work.

\subsection{Transverse Ising-model}

In order to show the capability of dealing with fermionic as well as non-particle-preserving Hamiltonians we consider the Ising model as second example.
More precisely, we consider a chain of spins subject to a transverse magnetic field which can be modeled by the Hamiltonian
\begin{align}
	\hat{H}_{\sigma} &= \sum_{i = 1}^{N} \sigma_{i}^{z} + \lambda \sum_{i = 1}^{N-1} \sigma_{i}^{x} \sigma_{i+1}^{x}.
	\label{Eq:transverse_ising_spin_hamiltonian}
\end{align}
Here, $\lambda$ denotes the ratio between the magnetic field strength in $z$-direction and the nearest-neighbor coupling strength in $x$-direction and $\sigma_{i}^{z}$ and $\sigma_{i}^{x}$ denote the Pauli matrices acting on spin $i$ defined as
\begin{align}
	\sigma_{i}^z &:= 
	\begin{pmatrix}
		1 & 0 \\
		0 & -1
	\end{pmatrix},
	&
	\sigma_{i}^x &:= 
	\begin{pmatrix}
		0 & 1 \\
		1 & 0
	\end{pmatrix}.
\end{align}
In particular, we set $\lambda = 1.2$ in our example.

The spin Hamiltonian in \cref{Eq:transverse_ising_spin_hamiltonian} is already perfectly suited for preparing thermal states via TEBD.
However, in order to make this model accessible for our preparation scheme we exploit a well-known result which states that the Ising model can be phrased equivalently in terms of fermionic operators $\hat{f}_i$ and $\hat{f}_i^\dagger$ \cite{Nielsen2005,Parkinson2010}.
This is an immediate consequence of applying the Jordan-Wigner transformation~\cite{Jordan1928}, described in more detail in \cref{Sec:jordan_wigner_transformation}, to the Hamiltonian in \cref{Eq:transverse_ising_spin_hamiltonian}.
In particular, we find
\begin{align}
	\sigma_{i}^{z} &= \hat{f}_i^\dagger \hat{f}_i - \hat{f}_i \hat{f}_i^\dagger
\end{align}
as well as
\begin{align}
	\sigma_{i}^{x} \sigma_{i+1}^{x} &= (\hat{f}_i^\dagger - \hat{f}_i) (\hat{f}_{i+1} + \hat{f}_{i+1}^\dagger) \\
	&= \hat{f}_{i}^\dagger \hat{f}_{i+1} + \hat{f}_{i}^\dagger \hat{f}_{i+1}^\dagger - \hat{f}_{i} \hat{f}_{i+1} - \hat{f}_{i} \hat{f}_{i+1}^\dagger. 
\end{align}
Hence, \cref{Eq:transverse_ising_spin_hamiltonian} maps to the active, fermionic Hamiltonian
\begin{align} 
	\begin{split}
		\hat{H}_f &= \sum_{i = 1}^{N} (\hat{f}_i^\dagger \hat{f}_i - \hat{f}_i \hat{f}_i^\dagger)  \\
		&+ \lambda \sum_{i = 1}^{N-1} (\hat{f}_{i}^\dagger \hat{f}_{i+1} + \hat{f}_{i}^\dagger \hat{f}_{i+1}^\dagger - \hat{f}_{i} \hat{f}_{i+1} - \hat{f}_{i} \hat{f}_{i+1}^\dagger).
	\end{split}
\end{align}
This Gaussian Hamiltonian serves as a starting point for our procedure to construct thermal states of the Ising model.

The results depicted in the bottom panel of \cref{Fig:fpos_vs_beta} are obtained similarly to the bosonic example. 
Thus, we again optimize over the various simulation parameters such that $\epsilon_{\bm{m}} < 10^{-2}$ and $\epsilon_{\gamma}^{\text{rel}} < 10^{-2}$ and the computational effort measured in terms of fpos becomes minimal.
Note that, in contrast to bosons, there is no need for a truncation of the Hilbert space for fermions due to Pauli's principle.
Hence, the number of parameters to optimize reduces by one.
\cref{Tab:ising_overview_beta_vs_fpos_gauco} and \cref{Tab:ising_overview_beta_vs_fpos_itime} summarize the different configurations considered and highlight the best configuration in boldface.
As in the bosonic benchmarks, a solution is only considered optimal if it meets the optimality criteria and is convergent.

\subsection{Alternative strategy for low temperatures}

For both examples, we note that at extremely low temperatures one also needs to assess an alternative strategy since in this case the thermal state is a superposition of very few low-energy states.
Hence, in principle, a thermal state could also be prepared by computing these first few excited states via DMRG and mixing them according to their thermal weights.
However, in the considered temperature range we were not able to construct the thermal state with the prescribed accuracy of the first and second moments with less than two excited states.
Hence, the preparation of the thermal state turns out to be a non-trivial task even for this strategy.

\onecolumngrid

\begin{table}[htbp]%[H]
	\begin{center}
		\caption{\textbf{Gaussian scheme optimization parameters for $\beta_{\text{resc}}$ vs. fpos and spin-boson model:}
		The table summarizes the configurations considered in the optimization procedure leading to the results depicted in \cref{Fig:fpos_vs_beta}.
		We highlight the best configuration for each $\beta_{\text{resc}}$ in boldface and state the minimal fpo count in the last column.
		}
		\label{Tab:sbm_overview_beta_vs_fpos_gauco}
		\renewcommand{\arraystretch}{1.5}
		\begin{tabular}{c || c | c | c | c | c | c}
			$\beta_{\text{resc}}$ & $M$             & Decomposition           & $r_{\text{max}}^{\text{state}}$ & $\epsilon_{\text{rel}}^{\text{state}}$ & $\epsilon_{\text{rel}}^{\text{gate}}$ & fpos \\\hline
			$0.5$                 & \textbf{14}, 16 & \textbf{Reck}           & 40, \textbf{60}, 80             & $\mathbf{10^{-5}}$, $10^{-7}$          & -                                     & $1.79 \cdot 10^{13}$ \\\hline
			$1.0$                 & \textbf{8},  10 & \textbf{Reck}, Clements & 10, \textbf{20}, 40             & $\mathbf{10^{-5}}$, $10^{-7}$          & -                                     & $2.01 \cdot 10^{10}$\\\hline
			$2.0$                 & \textbf{4},   6 & \textbf{Reck}, Clements & 5, \textbf{10}, 20              & $\mathbf{10^{-5}}$, $10^{-7}$          & -                                     & $6.20 \cdot 10^{7}$ \\\hline
			$3.0$                 & \textbf{3},   4 & \textbf{Reck}, Clements & 3, \textbf{5}, 10               & $\mathbf{10^{-5}}$, $10^{-7}$          & -                                     & $6.76 \cdot 10^{6}$ \\\hline
			$4.0$                 & \textbf{3},   4 & \textbf{Reck}, Clements & 3, \textbf{5}, 10               & $\mathbf{10^{-5}}$, $10^{-7}$          & -                                     & $3.31 \cdot 10^{6}$ \\\hline
		\end{tabular}
	\end{center}
\end{table}

\begin{table}[htbp]%[H]
	\begin{center}
		\caption{\textbf{Imaginary time evolution optimization parameters for $\beta_{\text{resc}}$ vs. fpos and spin-boson model:}
		The table summarizes the configurations considered in the optimization procedure leading to the results depicted in \cref{Fig:fpos_vs_beta}.
		We highlight the best configuration for each $\beta_{\text{resc}}$ in boldface and state the minimal fpo count in the last column.
		}
		\label{Tab:sbm_overview_beta_vs_fpos_itime}
		\renewcommand{\arraystretch}{1.5}
		\begin{tabular}{c || c | c | c | c | c | c | c}
			$\beta_{\text{resc}}$ & $M$              & $r_{\text{max}}^{\text{state}}$ & $\epsilon_{\text{rel}}^{\text{state}}$ & $\epsilon_{\text{rel}}^{\text{gate}}$ & S.-T. order      & $\mathrm{d}t$                 & fpos\\\hline
			$0.5$                 & \textbf{14}, 16  & 40, \textbf{60}, 80             & $\mathbf{10^{-5}}$, $10^{-7}$          & $10^{-7}$                             & \textbf{1}, 2, 4 & $10^{-2}$, $\mathbf{10^{-3}}$ & $2.14 \cdot 10^{15}$\\\hline
			$1.0$                 & \textbf{8}, 10   & 20, \textbf{40}, 60             & $\mathbf{10^{-5}}$, $10^{-7}$          & $10^{-7}$                             & \textbf{1}, 2, 4 & $10^{-2}$, $\mathbf{10^{-3}}$ & $2.00 \cdot 10^{14}$\\\hline
			$2.0$                 & \textbf{4}, 6, 8 & 20, \textbf{40}, 60             & $\mathbf{10^{-5}}$, $10^{-7}$          & $10^{-7}$                             & 1, 2, \textbf{4} & $\mathbf{10^{-2}}$, $10^{-3}$ & $2.03 \cdot 10^{13}$\\\hline
			$3.0$                 & \textbf{3}, 4    & 10, \textbf{20}, 40             & $\mathbf{10^{-5}}$, $10^{-7}$          & $10^{-7}$                             & 1, 2, \textbf{4} & $\mathbf{10^{-2}}$, $10^{-3}$ & $8.31 \cdot 10^{11}$\\\hline
			$4.0$                 & \textbf{3}, 4    & 10, \textbf{20}, 40             & $\mathbf{10^{-5}}$, $10^{-7}$          & $10^{-7}$                             & 1, 2, \textbf{4} & $\mathbf{10^{-2}}$, $10^{-3}$ & $8.31 \cdot 10^{11}$\\\hline
		\end{tabular}
	\end{center}
\end{table}

\begin{table}[htbp]%[H]
	\begin{center}
		\caption{\textbf{Gaussian scheme optimization parameters for $N$ vs. fpos and spin-boson model:}
		The table summarizes the configurations considered in the optimization procedure leading to the results depicted in \cref{Fig:modes_vs_fpos}.
		We highlight the best configuration for each $\beta_{\text{resc}}$ in boldface and state the minimal fpo count in the last column.}
		\label{Tab:sbm_overview_modes_vs_fpos_gauco}
		\renewcommand{\arraystretch}{1.5}
		\begin{tabular}{c | c || c | c | c | c | c | c}
			$\beta_{\text{resc}} = 1.0$ & $N$ & $M$            & Decomposition           & $r_{\text{max}}^{\text{state}}$ & $\epsilon_{\text{rel}}^{\text{state}}$ & $\epsilon_{\text{rel}}^{\text{gate}}$ & fpos\\\hline\hline
			                            & 10  & \textbf{8}, 10 & \textbf{Reck}, Clements &  10, \textbf{20}, 40            & $\mathbf{10^{-5}}$, $10^{-7}$          & -                                     & $7.95 \cdot 10^{9}$\\\hline
			                            & 20  & \textbf{8}, 10 & \textbf{Reck}, Clements &  10, \textbf{20}, 40            & $\mathbf{10^{-5}}$, $10^{-7}$          & -                                     & $2.01 \cdot 10^{10}$\\\hline
			                            & 30  & \textbf{8}, 10 & \textbf{Reck}, Clements &  10, \textbf{20}, 40            & $\mathbf{10^{-5}}$, $10^{-7}$          & -                                     & $4.18 \cdot 10^{10}$\\\hline
			                            & 40  & \textbf{8}, 10 & \textbf{Reck}, Clements &  10, \textbf{20}, 40            & $\mathbf{10^{-5}}$, $10^{-7}$          & -                                     & $8.65 \cdot 10^{10}$\\\hline
			                            & 80  & \textbf{8}, 10 & \textbf{Reck}, Clements &  10, \textbf{20}, 40            & $\mathbf{10^{-5}}$, $10^{-7}$          & -                                     & $1.86 \cdot 10^{11}$\\\hline
			$\beta_{\text{resc}} = 2.0$ & $N$ & $M$            & Decomposition           & $r_{\text{max}}^{\text{state}}$ & $\epsilon_{\text{rel}}^{\text{state}}$ & $\epsilon_{\text{rel}}^{\text{gate}}$ & fpos\\\hline\hline
			                            & 10  & \textbf{4}, 6  & \textbf{Reck}, Clements &  5, \textbf{10}, 20             & $\mathbf{10^{-5}}$, $10^{-7}$          & -                                     & $2.11 \cdot 10^{7}$\\\hline
			                            & 20  & \textbf{4}, 6  & \textbf{Reck}, Clements &  5, \textbf{10}, 20             & $\mathbf{10^{-5}}$, $10^{-7}$          & -                                     & $6.20 \cdot 10^{7}$\\\hline
			                            & 30  & \textbf{4}, 6  & \textbf{Reck}, Clements &  5, \textbf{10}, 20             & $\mathbf{10^{-5}}$, $10^{-7}$          & -                                     & $1.90 \cdot 10^{8}$\\\hline
			                            & 40  & \textbf{4}, 6  & \textbf{Reck}, Clements &  5, \textbf{10}, 20             & $\mathbf{10^{-5}}$, $10^{-7}$          & -                                     & $5.90 \cdot 10^{8}$\\\hline
			                            & 80  & \textbf{4}, 6  & \textbf{Reck}, Clements &  5, \textbf{10}, 20             & $\mathbf{10^{-5}}$, $10^{-7}$          & -                                     & $1.91 \cdot 10^{9}$\\\hline
			$\beta_{\text{resc}} = 3.0$ & $N$ & $M$            & Decomposition           & $r_{\text{max}}^{\text{state}}$ & $\epsilon_{\text{rel}}^{\text{state}}$ & $\epsilon_{\text{rel}}^{\text{gate}}$ & fpos\\\hline\hline
			                            & 10  & \textbf{3}, 4  & \textbf{Reck}, Clements &  3, \textbf{5}, 10              & $\mathbf{10^{-5}}$, $10^{-7}$          & -                                     & $2.34 \cdot 10^{6}$\\\hline
			                            & 20  & \textbf{3}, 4  & \textbf{Reck}, Clements &  3, \textbf{5}, 10              & $\mathbf{10^{-5}}$, $10^{-7}$          & -                                     & $6.76 \cdot 10^{6}$\\\hline
			                            & 30  & \textbf{3}, 4  & \textbf{Reck}, Clements &  3, \textbf{5}, 10              & $\mathbf{10^{-5}}$, $10^{-7}$          & -                                     & $1.37 \cdot 10^{7}$\\\hline
			                            & 40  & \textbf{3}, 4  & \textbf{Reck}, Clements &  3, \textbf{5}, 10              & $\mathbf{10^{-5}}$, $10^{-7}$          & -                                     & $5.19 \cdot 10^{7}$\\\hline
			                            & 80  & \textbf{3}, 4  & \textbf{Reck}, Clements &  3, \textbf{5}, 10              & $\mathbf{10^{-5}}$, $10^{-7}$          & -                                     & $1.94 \cdot 10^{8}$\\\hline
		\end{tabular}
	\end{center}
\end{table}

\begin{table}[htbp]%[H]
	\begin{center}
		\caption{\textbf{Gaussian scheme optimization parameters for $\beta_{\text{resc}}$ vs. fpos and Ising model:}
		The table summarizes the configurations considered in the optimization procedure leading to the results depicted in \cref{Fig:fpos_vs_beta}.
		We highlight the best configuration for each $\beta_{\text{resc}}$ in boldface and state the minimal fpo count in the last column.
		}
		\label{Tab:ising_overview_beta_vs_fpos_gauco}
		\renewcommand{\arraystretch}{1.5}
		\begin{tabular}{c || c | c | c | c | c}
			$\beta_{\text{resc}}$              & Decomposition                   & $r_{\text{max}}^{\text{state}}$ & $\epsilon_{\text{rel}}^{\text{state}}$ & $\epsilon_{\text{rel}}^{\text{gate}}$ & fpos \\\hline
			$0.01$   & \textbf{Reck}, Clements &  10, \textbf{20}, 40            & $\mathbf{10^{-5}}$, $10^{-7}$          & -                                     & $7.77 \cdot 10^{8}$ \\\hline
			$0.1$    & \textbf{Reck}, Clements &  20, \textbf{40}, 60            & $\mathbf{10^{-5}}$, $10^{-7}$          & -                                     & $2.20 \cdot 10^{9}$\\\hline
			$1.0$    & \textbf{Reck}, Clements &  20, \textbf{40}, 60            & $\mathbf{10^{-5}}$, $10^{-7}$          & -                                     & $1.86 \cdot 10^{9}$ \\\hline
			$10.0$   & \textbf{Reck}, Clements &  20, \textbf{40}, 60            & $\mathbf{10^{-5}}$, $10^{-7}$          & -                                     & $1.63 \cdot 10^{9}$ \\\hline
			$100.0$  & \textbf{Reck}, Clements &  20, \textbf{40}, 60            & $\mathbf{10^{-5}}$, $10^{-7}$          & -                                     & $1.62 \cdot 10^{9}$ \\\hline
		\end{tabular}
	\end{center}
\end{table}

\begin{table}[htbp]%[H]
	\begin{center}
		\caption{\textbf{Imaginary time evolution optimization parameters for $\beta_{\text{resc}}$ vs. fpos and Ising model:}
		The table summarizes the configurations considered in the optimization procedure leading to the results depicted in \cref{Fig:fpos_vs_beta}.
		We highlight the best configuration for each $\beta_{\text{resc}}$ in boldface and state the minimal fpo count in the last column.
		}
		\label{Tab:ising_overview_beta_vs_fpos_itime}
		\renewcommand{\arraystretch}{1.5}
		\begin{tabular}{c || c | c | c | c | c | c}
			$\beta_{\text{resc}}$  & $r_{\text{max}}^{\text{state}}$ & $\epsilon_{\text{rel}}^{\text{state}}$ & $\epsilon_{\text{rel}}^{\text{gate}}$ & S.-T. order      & $\mathrm{d}t$                  & fpos\\\hline
			$0.01$                 & \textbf{5}, 10, 20, 40          & $\mathbf{10^{-5}}$, $10^{-7}$          & $10^{-7}$                             & 1, \textbf{2}, 4 & $\mathbf{10^{-2}}$, $10^{-3}$           & $7.41 \cdot 10^{5}$\\\hline
			$0.1$                  & 5, \textbf{10}, 20, 40          & $\mathbf{10^{-5}}$, $10^{-7}$          & $10^{-7}$                             & \textbf{1}, 2, 4 & $10^{-1}$,$10^{-2}$, $\mathbf{10^{-3}}$ & $5.07 \cdot 10^{8}$\\\hline
			$1.0$                  & 5, 10, \textbf{20}, 40          & $\mathbf{10^{-5}}$, $10^{-7}$          & $10^{-7}$                             & \textbf{1}, 2, 4 & $10^{-1}$,$10^{-2}$, $\mathbf{10^{-3}}$ & $4.03 \cdot 10^{10}$\\\hline
			$10.0$                 & 5, 10, \textbf{20}, 40          & $\mathbf{10^{-5}}$, $10^{-7}$          & $10^{-7}$                             & \textbf{1}, 2, 4 & $10^{-1}$,$10^{-2}$, $\mathbf{10^{-3}}$ & $4.24 \cdot 10^{11}$\\\hline
			$100.0$                & 5, 10, \textbf{20}, 40          & $\mathbf{10^{-5}}$, $10^{-7}$          & $10^{-7}$                             & \textbf{1}, 2, 4 & $10^{-1}$,$10^{-2}$, $\mathbf{10^{-3}}$ & $4.26 \cdot 10^{12}$\\\hline
		\end{tabular}
	\end{center}
\end{table}

\twocolumngrid

\end{document}